\documentclass[sn-vancouver,Numbered]{sn-jnl}% Math and Physical Sciences Reference Style

\usepackage{graphicx}%
\usepackage{multirow}%
\usepackage{amsmath,amssymb,amsfonts}%
\usepackage{amsthm}%
\usepackage{mathrsfs}%
\usepackage{float}
\usepackage[title]{appendix}%
\graphicspath{{./Figures/}}
\usepackage{subcaption}
\usepackage{xcolor}%
\usepackage{tabularx}
\usepackage{tabularray}
\usepackage{qcircuit}
\usepackage{textcomp}%
\usepackage{manyfoot}%
\usepackage{braket}
\usepackage{colortbl}
\usepackage{dcolumn}
\usepackage{booktabs}%
\usepackage{algorithm}
\usepackage{algorithmicx}%
\usepackage{algpseudocode}
\usepackage{listings}%
\begin{document}

\title[Article Title]{Quantum-Inspired Geometric Classification with Correlation Group Structures and VQC Decision Modeling}

\author{\fnm{Arya Ansuman} \sur{Priyadarshi}}\email{aryapriyadarshi@icloud.com}

\author*[1]{\fnm{Nishikanta} \sur{Mohanty}}\email{nishikanta2000@gmx.com}

\author[2,3]{\fnm{Bikash K.} \sur{Behera}}\email{bikas.riki@gmail.com}

\author[4]{\fnm{Badshah} \sur{Mukherjee}}\email{badshah.mukherjee@outlook.com}

\affil[1]{\orgdiv{Centre for Quantum Software and Information}, \orgname{University of Technology Sydney}, \orgaddress{\street{15 Broadway, Ultimo}, \city{Sydney}, \postcode{2007}, \state{NSW}, \country{Australia}}}

\affil[2]{\orgdiv{} \orgname{Bikash's Quantum (OPC) Pvt. Ltd.}, \orgaddress{\street{Balindi}, \city{Mohanpur}, \postcode{741246}, \state{WB}, \country{India}}}
\affil[3]{\orgname{Università degli Studi di Cagliari}, \orgaddress{\street{Via Is Mirrions}, \city{Cagliari}, \postcode{09123}, \country{Italy}}}

%\affil[3]{\orgdiv{} \orgname{}, \orgaddress{ \city{Dubai}, \postcode{9262}, \country{UAE}}}

\abstract{We propose a geometry-driven quantum-inspired classification framework that integrates Correlation Group Structures (CGR), compact SWAP-test-based overlap estimation, and selective variational quantum decision modeling. Rather than directly approximating class posteriors, the method adopts a geometry-first paradigm in which samples are evaluated relative to class medoids using overlap-derived Euclidean-like and angular similarity channels. CGR organizes features into anchor-centered correlation neighborhoods, generating nonlinear, correlation-weighted representations that enhance robustness in heterogeneous tabular spaces. These geometric signals are fused through a non-probabilistic margin-based fusion score, serving as a lightweight and data-efficient primary classifier for small-to-moderate datasets. On Heart Disease, Breast Cancer, and Wine Quality datasets, the fusion-score classifier achieves 0.8478, 0.8881, and 0.9556 test accuracy respectively, with macro-F1 scores of 0.8463, 0.8703, and 0.9522, demonstrating competitive and stable performance relative to classical baselines. For large-scale and highly imbalanced regimes, we construct compact $\Delta$-distance contrastive features and train a variational quantum classifier (VQC) as a nonlinear refinement layer. On the Credit Card Fraud dataset (0.17\% prevalence), the $\Delta$+VQC pipeline achieves $\approx$0.85 minority recall at an alert rate of $\approx$1.31\%, with ROC-AUC 0.9249 and PR-AUC 0.3251 under full-dataset evaluation. These results highlight the importance of operating-point-aware assessment in rare-event detection and demonstrate that the proposed hybrid geometric–variational framework provides interpretable, scalable, and regime-adaptive classification across heterogeneous data settings.}

\keywords{Compact Swap Test, Correlation Group Structures,  Medoids, Coordinate Descent Optimisation, VQC Decision Modeling, Quantum Classification}

\maketitle

\section{Introduction}

Modern machine learning continues to face persistent challenges when operating within high-dimensional feature spaces and heterogeneous data regimes. In domains such as fraud detection, medical diagnostics, and risk screening, classification performance is frequently constrained by limited labeling, extreme class imbalance, and non-stationary (concept-drifting) distributions \cite{HeGarcia2009,Gama2014}. Traditional supervised models often struggle when the decision boundary is governed by a complex mixture of global structure and localized sub-population patterns. These issues are exacerbated in high dimensions, where distance concentration and the broader ``curse of dimensionality'' can erode the discriminative contrast of standard similarity measures and nearest-neighbor style reasoning \cite{Beyer1999,Pestov1999}. A robust alternative is \emph{prototype-driven similarity representation}, where samples are assessed based on their relationship to class representatives (e.g., medoids). Medoid-based clustering and classification are well established due to their interpretability and robustness properties, particularly compared to mean-based prototypes in the presence of outliers \cite{KaufmanRousseeuw1987}. However, the efficacy of prototype-based approaches depends critically on the discriminative power of the similarity metrics used. In high-dimensional settings, standard metrics may become less informative and can conflate magnitude and directional components, motivating hybrid geometric representations that explicitly separate distance-like and angle-like structure \cite{Beyer1999}.

To address this, we leverage \emph{quantum-inspired primitives}, specifically SWAP-test-style overlap estimation, to compute similarity signals via interference-based routines \cite{Buhrman2001}. Under amplitude-style encodings, overlap estimation provides a principled mechanism to access inner-product structure within a high-dimensional feature space using a number of qubits that scales logarithmically with the feature dimension (in the state representation), enabling the construction of geometry-aware similarity features \cite{Havlicek2019,SchuldKilloran2019}. By mapping overlaps into distinct angular and distance-like components, we obtain a hybrid representation that captures complementary geometric information often missed by purely classical measures. Similarity signals alone, however, can be unstable when latent sub-population structure is ignored, particularly in imbalanced regimes where majority-class geometry can dominate the similarity landscape \cite{HeGarcia2009,Chawla2002}. We therefore propose a \emph{Correlation Group Structure} mechanism a Cluster/Group-Responsive (CGR) construction to partition data into structured configurations that make similarity aggregation responsive to underlying clusters. Over this CGR configuration space, we apply coordinate-descent-driven model selection to identify the best-performing group-combination and calibration parameters for a target metric. Once the optimal CGR structure is fixed, we compute CGR-conditioned distance--angle deltas and train a lightweight variational quantum classifier (VQC) as the final decision model, aligning with established variational/quantum-feature-space learning paradigms \cite{Havlicek2019,SchuldKilloran2019}.

\section{Quantum Distance Classification}
\paragraph{Geometric classification versus probabilistic classification.}
Most supervised classification pipelines are presented in a probabilistic form: given an input $x$, a model produces class scores that are interpreted as an estimate of a posterior distribution $p(y\mid x)$ (e.g., logistic regression, neural networks, and many calibrated ensemble methods). Decision boundaries in such models are typically learned by minimizing a likelihood-based objective (or a surrogate such as cross-entropy), and the learned parameters may not retain a direct geometric interpretation in the original feature space. By contrast, \emph{distance-based} or \emph{geometric} classification assigns labels based on the position of $x$ relative to representative structures in the data (e.g., prototypes, medoids, centroids, or sub-manifolds). In its simplest form, a geometric classifier selects the label of the nearest prototype under a chosen metric, or uses a margin defined by the difference between distances to competing class representatives. This paradigm is often preferred when interpretability is required, when decision logic naturally depends on similarity (e.g., ``is this case closer to known fraud patterns or normal patterns?''), or when data are heterogeneous and local geometry is more stable than a global parametric surface.

\paragraph{Geometric Foundations of Quantum Classification}

A distinguishing advantage of quantum information processing is its role as a native toolbox for computing \emph{geometric relationships} between data representations. Unlike classical architectures that may require significant computational overhead to determine high-dimensional similarities, quantum states exist within a Hilbert space where inner products, overlaps, and fidelities are fundamental, physically measurable quantities. Consequently, quantum routines are often inherently designed to estimate similarity measures—either directly through interference mechanisms, such as SWAP-test-style circuits, or indirectly through the expectation values of observables. This physical architecture aligns seamlessly with \emph{geometric classification} paradigms, where decision boundaries are defined by distances, angles, margins, or kernel similarities. Within this framework, overlap estimation serves as a primitive that maps to distance-like and angular measures for normalized encodings. These can be further composed into prototype-based decision rules, effectively linking quantum measurements to interpretable geometric signals \cite{Buhrman2001, Havlicek2019, SchuldKilloran2019}.

\paragraph{SWAP test and overlap estimation.}
A canonical quantum primitive for comparing two quantum states is the SWAP test, which estimates the squared overlap
\begin{equation}
s(x,x') \;=\; \left|\langle \psi(x)\,|\,\psi(x')\rangle\right|^2,
\label{eq:swap_overlap_general}
\end{equation}
via a simple interference circuit and a measurement on an ancilla qubit \cite{Buhrman2001}. Overlap estimation underpins a range of quantum machine learning constructions, including quantum kernels and feature-space methods, where the overlap (or a function of it) plays the role of a similarity function in a Hilbert space \cite{Havlicek2019,SchuldKilloran2019}. In distance-based quantum classification, overlap estimates can be mapped into geometric quantities such as angular separation or Euclidean-like dissimilarities for normalized vectors. For example, one may define an angle
\begin{equation}
\Theta(x,x') \;=\; \arccos\!\left(\sqrt{s(x,x')}\right),
\label{eq:angle_from_overlap_general}
\end{equation}
and a monotone distance-like mapping
\begin{equation}
D(x,x') \;=\; \sqrt{2\left(1-\sqrt{s(x,x')}\right)},
\label{eq:distance_from_overlap_general}
\end{equation}

\paragraph{Recent Approaches to Quantum Geometric Classification}
Building on these primitives, several approaches have formalized quantum formulations of distance- and similarity-based learning. Early investigations into quantum nearest-neighbor acceleration demonstrated that, provided efficient state preparation is feasible, the ability to compute inner products can significantly enhance similarity search and related classification tasks \cite{Wiebe2015}. Subsequent proposals introduced explicit distance-based classifiers that utilize interference circuits to evaluate the proximity between a query state and class representatives, facilitating label assignment via nearest-prototype rules \cite{Schuld2017Distance}. More broadly, the field has seen the rise of quantum kernel methods, which define a similarity function $k(x,x') = |\langle \phi(x)|\phi(x')\rangle|^2$ within an implicit feature space. By employing classical learning algorithms on the resulting quantum-computed kernel matrix, these methods place overlap estimation at the center of the learning pipeline, bridging the gap between quantum state interference and established statistical learning theory \cite{Havlicek2019, SchuldKilloran2019}.

\section{Novelty and Contribution}\label{Contribution}
In this work, we propose a geometric classification paradigm and use compact SWAP-test-based overlap estimation as the primary mechanism to compute similarity signals between samples and representative prototypes. Unlike purely probabilistic classifiers that directly learn $p(y\mid x)$ from raw features, our approach first constructs geometry-aware similarity features including both distance-like and angle-like components derived from overlaps and organizes them under correlation group structures (CGR) to improve robustness in heterogeneous feature regimes.

Crucially, the proposed framework is \emph{hybrid and data-regime-aware}. We introduce a \emph{Fusion Score} as the primary, lightweight classifier that aggregates CGR group-wise geometric evidence into class scores.Fusion Score is a \emph{non-probabilistic} classifier: it produces per-class margin-like dissimilarity scores from overlap-derived geometric evidence and predicts via an $\arg\min$ decision rule, without explicitly modeling calibrated posteriors $p(y\mid x)$. Empirically, Fusion Score is highly effective on small-to-moderate datasets (Heart Disease, Wine Quality, Breast Cancer), where the class structure is well captured by prototype distances and a parametric learner may struggle due to limited sample support. However, in large, highly imbalanced and behaviourally complex datasets (Fraud), Fusion Score alone can underfit because discrimination depends on non-linear interactions across groups and subtle margin patterns.

To address this, we construct $\Delta$-distance margin features from the CGR distance tensor and train a variational quantum classifier (VQC) on these $\Delta$-features as a \emph{selective refinement} (or residual decision model). In practice, the VQC is invoked when Fusion Score is not optimal, for example, under explicit imbalance/complexity regimes, or for samples exhibiting low fusion confidence (small best-vs-runner-up margin). This design retains interpretability and stability from geometric fusion in low-data settings, while enabling non-linear correction via VQC where large-scale, imbalanced structure demands a richer decision boundary. The overall process is divided into the following steps after initial data preparation and preprocessing:

\begin{itemize}
    \item Correlation Group structure formation.
    \item Euclidean Medoid calculation as per available classes
    \item Quantum distance calculation (Euclidean and Angular) by Compact Swaptest.
    \item Finding the optimal configuration by Coordinate descent Optimisation
    \item Measuring  the classification performance by Fusion score (angular score and Euclidean score)
    \item Calculating distance deltas based on the Optimal configuration
    \item Train VQC on distance deltas as final classifier.
\end{itemize}

The detailed implementation of these steps, including prototype construction, correlation group configuration, coordinate-descent model selection, and VQC training, is presented in Section~\ref{Algorithms}.

\section{Process Flow and algorithms}\label{Algorithms}
This section presents the end-to-end methodology of the proposed classification process. The central objective of the framework is to construct a robust, geometry-driven decision pipeline in which quantum overlap estimation is used as a primary similarity primitive, while correlation-aware grouping and coordinate-descent selection ensure stability across heterogeneous data regimes. Unlike conventional probabilistic classifiers that learn a direct mapping from raw features to posteriors, our approach explicitly decomposes classification into (i) representation by \emph{prototype-referenced geometry} and (ii) decision modeling on a compact set of \emph{contrastive delta features}. This design yields interpretability (via distances, angles, and margins), configurability (via group-structure selection), and a clear separation between similarity extraction and final decision learning. At a high level, the pipeline begins by forming \emph{Correlation Group Structures} (CGR) that capture latent sub-population patterns and define a discrete configuration space of feasible group-combination ``configs.'' Within each config, class representatives are computed using Euclidean medoids, providing robust prototypes that are less sensitive to outliers than mean-based representatives. We then compute quantum-inspired geometric similarities between each sample and the medoids using a \emph{compact SWAP-test} routine, which estimates state overlaps and maps them into two complementary channels: a distance-like (Euclidean) measure and an angular measure. These channels provide a richer geometric representation than either metric alone, particularly in high-dimensional settings where magnitude- and direction-based separations may differ.

Given the CGR configuration space, we apply a coordinate-descent optimization procedure to identify the \emph{optimal configuration} for a target performance metric. This stage simultaneously calibrates the relative contribution of Euclidean and angular similarity channels and selects the configuration that yields the best empirical performance under the chosen evaluation criterion. Classification performance is first measured through a \emph{fusion score}, which combines the Euclidean and angular signals into a margin-like decision statistic and serves as a strong, lightweight geometric classifier in its own right. After selecting the optimal configuration, we compute \emph{distance deltas} contrastive differences between class-referenced distances/angles under the selected CGR structure to obtain a compact and stable feature vector per sample. Finally, we train a variational quantum classifier (VQC) on these delta features as a \emph{refinement} decision model, particularly suited to regimes where fusion scores alone are not optimal (e.g., large, highly imbalanced, and behaviourally complex datasets). In contrast, for small-to-moderate datasets where sample support is limited, the fusion score often provides the most reliable performance while a VQC may be data-constrained. The methodology is formalized in the remainder of this section through a sequence of algorithms and their associated mathematical definitions, presented in the order as per Section~\ref{Contribution}.

%%\subsection{Algorithm 1: Create Correlation Group Structures}
\subsection{Correlation Group Structures and Nonlinear Feature Construction}
\label{subsec:cgr_nonlinear}

This subsection formally defines \emph{Correlation Group Structures} (CGR) and introduces the nonlinear feature construction used throughout the proposed framework. The CGR mechanism partitions the original feature space into anchor-centered correlation neighborhoods and then induces nonlinear features by combining (i) correlation-weighted feature groups and (ii) correlation group configurations. This yields a compact representation that preserves local dependency structure while enabling expressive, interaction-like behavior without requiring explicit polynomial expansion. Let $X\in\mathbb{R}^{N\times M}$ denote the dataset with feature index set $\mathcal{F}=\{1,\dots,M\}$, and let $R\in[-1,1]^{M\times M}$ be the feature--feature correlation matrix computed on $X$ (excluding the target). The Correlation Group Structure (CGR) mechanism constructs anchor-specific correlation neighborhoods and converts them into nonlinear, correlation-structured features.

\paragraph{Correlation group membership.}
Each feature can act as an \emph{anchor} $a\in\mathcal{F}$. For a fixed group size parameter $m$, we define the membership set of anchor $a$ by ranking all other features by absolute correlation magnitude and selecting the top-$m$:
\begin{equation}
\mathcal{M}_a \;=\; \{a\}\cup \mathrm{TopM}\Big(\{|R_{a,f}|\}_{f\in \mathcal{F}\setminus\{a\}},\, m\Big).
\label{eq:cgr_membership_unified}
\end{equation}
The ranking in \eqref{eq:cgr_membership_unified} is sign-agnostic; the sign of $R_{a,f}$ may still be retained as a weight if required.

\paragraph{Anchor-wise nonlinear CGR feature (correlation-weighted $\ell_2$ aggregation).}
Given a record $x\in\mathbb{R}^M$, we define the correlation-weighted subvector induced by anchor $a$ as
\begin{equation}
v_a(x) \;=\; \big(R_{a,f}\,x_f\big)_{f\in\mathcal{M}_a}\in\mathbb{R}^{|\mathcal{M}_a|},
\label{eq:va_def_unified}
\end{equation}
and compute the anchor-wise CGR feature as the $\ell_2$ norm of this subvector:
\begin{equation}
\phi_a(x) \;=\; \|v_a(x)\|_2
\;=\;
\sqrt{\sum_{f\in\mathcal{M}_a}\big(R_{a,f}\,x_f\big)^2}.
\label{eq:phi_a_unified}
\end{equation}
Equation~\eqref{eq:phi_a_unified} is nonlinear in the original features due to the squaring and square-root operations, and it summarizes the joint magnitude of a correlation neighborhood around the anchor. If correlation weighting is not used in the aggregation, the unweighted alternative is
\begin{equation}
\phi_a(x) \;=\; \|x_{\mathcal{M}_a}\|_2
\;=\;
\sqrt{\sum_{f\in \mathcal{M}_a} x_f^2}.
\label{eq:phi_unweighted_unified}
\end{equation}

\paragraph{CGR embedding.}
Stacking the anchor-wise features yields the CGR representation
\begin{equation}
\Phi(x) \;=\; \big[\phi_1(x),\phi_2(x),\dots,\phi_M(x)\big]^\top \in \mathbb{R}^{M},
\qquad
\Phi(X)\in\mathbb{R}^{N\times M}.
\label{eq:cgr_embedding_unified}
\end{equation}

\paragraph{CGR-data multiplicative feature (correlation strength $\times$ group activation).}
In addition to the direct aggregation in \eqref{eq:phi_a_unified}--\eqref{eq:phi_unweighted_unified}, we also define a multiplicative nonlinear feature that explicitly separates correlation structure from record-level activation. First, we compute a dataset-level correlation-group strength for each anchor:
\begin{equation}
g_a \;=\; \left\|\big(|R_{a,f}|\big)_{f\in\mathcal{M}_a}\right\|_2
\;=\;
\sqrt{\sum_{f\in\mathcal{M}_a}|R_{a,f}|^2}.
\label{eq:cgr_strength_unified}
\end{equation}
Next, we compute the record-specific activation of the same variable set:
\begin{equation}
h_a(x) \;=\; \|x_{\mathcal{M}_a}\|_2
\;=\;
\sqrt{\sum_{f\in\mathcal{M}_a}x_f^2}.
\label{eq:data_energy_unified}
\end{equation}
The multiplicative CGR feature is then
\begin{equation}
z_a(x) \;=\; g_a\cdot h_a(x),
\label{eq:mult_feature_unified}
\end{equation}
and stacking across anchors produces $Z(x)=[z_1(x),\dots,z_M(x)]^\top$ and $Z(X)\in\mathbb{R}^{N\times M}$. The transformation $\phi_a(x)$ in \eqref{eq:phi_a_unified} and the product feature $z_a(x)$ in \eqref{eq:mult_feature_unified} both introduce nonlinearity through $\ell_2$ aggregation and multiplicative coupling. The term $g_a$ is constant for a fixed dataset (derived from $R$), whereas $h_a(x)$ varies across records, so $z_a(x)$ increases for records that strongly express a feature group that is also strongly correlated around anchor $a$. It presents the algorithms used to (i) construct CGR anchor memberships from a correlation matrix and (ii) generate a CGR feature matrix for any chosen configuration of anchor-specific subsets.

\begin{algorithm}[t]
\caption{Build CGR Anchors, Membership Sets, and Correlation Weights}
\label{alg:build_anchor_membership}
\begin{algorithmic}[1]
\Require Correlation matrix $R$ over features (and optionally target) with column set $\mathcal{F}\cup\{y\}$; target name $y$; top-$m$ parameter $m$
\Ensure Ordered anchors $\mathcal{A}$; membership map $\{\mathcal{M}_a\}_{a\in\mathcal{A}}$; weight map $\{\rho_{a,f}\}$
\vspace{2pt}

\State $\mathcal{F} \gets$ all columns in $R$ excluding $y$
\If{$y$ exists in $R$}
    \State $t_f \gets |R_{f,y}|$ for all $f\in\mathcal{F}$
    \State $\mathcal{A} \gets$ features sorted by $t_f$ in descending order
\Else
    \State $\mathcal{A} \gets \mathcal{F}$ \Comment{fallback ordering}
\EndIf

\For{each anchor $a\in\mathcal{A}$}
    \State $c_f \gets R_{a,f}$ for all $f\in\mathcal{F}$ \Comment{exclude target from membership}
    \State Sort features by $|c_f|$ descending
    \State $\widetilde{\mathcal{M}} \gets$ first $m$ features in the sorted list
    \State $\mathcal{M}_a \gets [a] \oplus [f\in\widetilde{\mathcal{M}}: f\neq a]$ \Comment{anchor first}
    \For{each member $f\in\mathcal{M}_a$}
        \If{$f=a$}
            \State $\rho_{a,f} \gets 1.0$
        \Else
            \State $\rho_{a,f} \gets R_{a,f}$ \Comment{if missing/NaN then set to $0$}
        \EndIf
    \EndFor
\EndFor
\State \Return $\mathcal{A}$, $\{\mathcal{M}_a\}$, $\{\rho_{a,f}\}$
\end{algorithmic}
\end{algorithm}

\begin{algorithm}[t]
\caption{Compute Anchor Feature Vector for a Given Anchor and Subset}
\label{alg:compute_anchor_feature_vector}
\begin{algorithmic}[1]
\Require Dataset $X\in\mathbb{R}^{N\times M}$ with feature columns; anchor $a$; subset $\kappa(a)\subseteq\mathcal{F}$; weights $\rho_{a,f}$
\Ensure Anchor feature vector $u_a \in \mathbb{R}^{N}$
\vspace{2pt}

\For{$i=1$ to $N$}
    \State $u_a(i) \gets \sqrt{\sum_{f\in\kappa(a)}\left(\rho_{a,f}\,X_{i,f}\right)^2}$
\EndFor
\State \Return $u_a$
\end{algorithmic}
\end{algorithm}

\begin{algorithm}[t]
\caption{Build CGR Feature Matrix for a Configuration}
\label{alg:build_feature_matrix_from_config}
\begin{algorithmic}[1]
\Require Dataset $X\in\mathbb{R}^{N\times M}$; anchors $\mathcal{A}=\{a_1,\dots,a_K\}$; configuration $\kappa$ with subsets $\kappa(a)$; weights $\rho_{a,f}$
\Ensure CGR feature matrix $F\in\mathbb{R}^{N\times K}$
\vspace{2pt}

\State Initialize $F \gets \mathbf{0}_{N\times K}$
\For{$j=1$ to $K$}
    \State $a \gets a_j$
    \State $F_{:,j} \gets \mathrm{AnchorFeatureVector}(X, a, \kappa(a), \rho)$ \Comment{Algorithm~\ref{alg:compute_anchor_feature_vector}}
\EndFor
\State \Return $F$
\end{algorithmic}
\end{algorithm}

\subsubsection{Class-Wise Medoid Computation on CGR-Induced Nonlinear Features}
\label{subsubsec:medoid_cgr_features}

After constructing the CGR-induced nonlinear representation, we compute class-wise medoids to obtain robust prototype representatives for distance-based classification. Medoids are preferred over mean-based prototypes because they are selected from observed samples and are therefore less sensitive to outliers and heavy-tailed feature distributions. %\paragraph{Nonlinear feature space.}
Let $Z(X)\in\mathbb{R}^{N\times d}$ denote the CGR-induced nonlinear feature matrix, where each record is mapped to a vector
\begin{equation}
z_i \;=\; Z(x_i)\in\mathbb{R}^{d}, \qquad i=1,\dots,N,
\label{eq:z_i_def}
\end{equation}
and $d$ is the embedding dimension (typically $d=M$ when using one anchor per original feature, but may be smaller if anchors are subsampled). %\paragraph{Class partitions.}
For each class $c\in\{0,1,\dots,C-1\}$, define the index set of samples in that class as
\begin{equation}
\mathcal{I}_c \;=\; \{\, i \in \{1,\dots,N\} \;:\; y_i = c \,\},
\qquad
N_c = |\mathcal{I}_c|.
\label{eq:class_index_set}
\end{equation}
Let $\mathcal{Z}_c = \{z_i : i\in\mathcal{I}_c\}$ denote the set of nonlinear feature vectors belonging to class $c$.

\paragraph{Euclidean medoid objective.}
The medoid of class $c$ is defined as the sample (in nonlinear feature space) that minimizes the total within-class Euclidean dissimilarity:
\begin{equation}
m_c \;=\; z_{j^*_c},
\qquad
j^*_c \;=\;
\arg\min_{j\in\mathcal{I}_c}\;
\sum_{i\in\mathcal{I}_c}
\left\| z_i - z_j \right\|_2.
\label{eq:medoid_objective}
\end{equation}
The vector $m_c\in\mathbb{R}^{d}$ serves as the prototype representative for class $c$ in subsequent quantum similarity computations. If a class exhibits multi-modal structure, the framework can be extended to $K$-medoids per class by selecting a set $\{m_{c,1},\dots,m_{c,K}\}$ that minimizes the standard $K$-medoids objective. The remainder of the pipeline remains unchanged, with quantum similarity computed against each medoid and aggregated according to the selected configuration. All medoid computations are performed \emph{after} transforming the original data into the CGR-induced nonlinear space. This ensures that class prototypes reflect correlation-structured group activations rather than raw feature coordinates, improving robustness in heterogeneous datasets where local dependency structure is predictive of class membership.

\begin{algorithm}[t]
\caption{Approximate Euclidean Medoid via Subsampling}
\label{alg:euclidean_medoid_subsample}
\begin{algorithmic}[1]
\Require Point set $P=\{p_i\}_{i=1}^{n}$, $p_i\in\mathbb{R}^{d}$; max subsample size $m_{\max}$; random seed $s$
\Ensure Approximate medoid $\mu \in \mathbb{R}^{d}$
\vspace{2pt}

\State $n \gets |P|$
\If{$n \le m_{\max}$}
    \State $S \gets P$
\Else
    \State Sample index set $\mathcal{J}\subset\{1,\dots,n\}$ with $|\mathcal{J}|=m_{\max}$ uniformly without replacement using seed $s$
    \State $S \gets \{p_j : j\in\mathcal{J}\}$
\EndIf

\State Compute pairwise distance matrix $\Delta \in \mathbb{R}^{|S|\times|S|}$ where $\Delta_{ij}=\|S_i - S_j\|_2$
\State Compute row-sums $u_i \gets \sum_{j=1}^{|S|}\Delta_{ij}$ for $i=1,\dots,|S|$
\State $i^\star \gets \arg\min_i u_i$
\State $\mu \gets S_{i^\star}$
\State \Return $\mu$
\end{algorithmic}
\end{algorithm}

\begin{algorithm}[t]
\caption{Fit Class-Wise Medoids in CGR Feature Space}
\label{alg:fit_class_medoids}
\begin{algorithmic}[1]
\Require Training feature matrix $F\in\mathbb{R}^{N\times d}$ (CGR-induced nonlinear features); labels $y\in\{0,\dots,C-1\}^N$; $m_{\max}$; seed $s$
\Ensure Medoid dictionary $\{\mu_c\}_{c\in\mathcal{C}}$
\vspace{2pt}

\State $\mathcal{C} \gets \mathrm{unique}(y)$
\For{each class $c\in\mathcal{C}$}
    \State $F_c \gets \{F_i : y_i=c\}$ \Comment{rows of $F$ belonging to class $c$}
    \State $\mu_c \gets \mathrm{EuclideanMedoidSubsample}(F_c, m_{\max}, s)$ \Comment{Algorithm~\ref{alg:euclidean_medoid_subsample}}
\EndFor
\State \Return $\{\mu_c\}_{c\in\mathcal{C}}$
\end{algorithmic}
\end{algorithm}

\subsubsection{Compact SWAP Test for Quantum Distance and Angular Similarity}
\label{subsubsec:compact_swaptest}

The compact SWAP-test routine used in this work is inherited from our Quantum-SMOTE feature-similarity module, where it was introduced as a resource-efficient overlap estimator for high-dimensional vectors under amplitude-style preparation. The same primitive applies directly to geometric classification because both settings require a stable estimate of the overlap between two data-derived quantum states. We retain the compact construction in which one state ($|\phi\rangle$) is prepared on a single qubit and the second state ($|\psi\rangle$) is prepared on a multi-qubit register. Compared to symmetric two-register SWAP-test implementations, this reduces circuit width and state-preparation overhead while preserving an ancilla-measurement statistic that maps to a squared-overlap estimate \cite{NielsenChuang2000,GarciaEscartin2013}. In this manuscript, the overlap estimate is transformed into two geometric signals: an angular distance (via cosine similarity) and a Euclidean distance reconstruction, which form the inputs to CGR configuration selection and downstream decision modeling. Given $x,y\in\mathbb{R}^{M}$, we compute (i) an angular distance capturing directional similarity and (ii) a Euclidean distance capturing magnitude separation. Under amplitude-style encoding, vectors are mapped to normalized quantum states and similarity is expressed through state overlaps \cite{SchuldPetruccione2018,SchuldKilloran2019}. If $\|x\|_2=0$ or $\|y\|_2=0$, state preparation is undefined and we use a classical fallback.

\paragraph{Degenerate handling (classical fallback).}
Let $n_x=\|x\|_2$ and $n_y=\|y\|_2$. If $n_x=0$ or $n_y=0$, we set
\begin{equation}
D(x,y)=\|x-y\|_2,
\label{eq:classical_fallback_distance}
\end{equation}
and
\begin{equation}
\Theta(x,y)=
\begin{cases}
\arccos\!\left(\dfrac{x^\top y}{\|x\|_2\,\|y\|_2}\right), & \|x\|_2>0,\ \|y\|_2>0,\\[8pt]
0, & \text{otherwise},
\end{cases}
\label{eq:classical_fallback_angle}
\end{equation}
with clipping of the cosine argument to $[-1,1]$ for numerical stability.

\paragraph{Compact state preparation.}
Assume $n_x>0$ and $n_y>0$ and define
\begin{equation}
Z = \|x\|_2^2 + \|y\|_2^2.
\label{eq:Z_def}
\end{equation}
We prepare a single-qubit state encoding relative norms,
\begin{equation}
|\phi\rangle
=
\frac{1}{\sqrt{Z}}
\left(
\|x\|_2\,|0\rangle
-
\|y\|_2\,|1\rangle
\right),
\label{eq:phi_state}
\end{equation}
and a multi-qubit state by interleaving normalized components,
\begin{equation}
|\psi\rangle
=
\frac{1}{\sqrt{2}}
\sum_{k=1}^{M}
\left(
\frac{x_k}{\|x\|_2}\,|2k-2\rangle
+
\frac{y_k}{\|y\|_2}\,|2k-1\rangle
\right),
\label{eq:psi_state}
\end{equation}
corresponding to an amplitude vector of length $2M$, padded to the nearest power-of-two for amplitude initialization \cite{SchuldPetruccione2018}.

\paragraph{Overlap estimate.}
The SWAP test estimates overlaps via ancilla-controlled interference and measurement \cite{NielsenChuang2000,GarciaEscartin2013}. Using ancilla $a$, a single qubit for $|\phi\rangle$, and a register for $|\psi\rangle$, we apply
\begin{equation}
H_a \;\rightarrow\; \mathrm{CSWAP}(a;\,\phi\text{-qubit},\,\psi\text{-qubit}_0)\;\rightarrow\; H_a,
\label{eq:compact_swap_skeleton}
\end{equation}
where the controlled-SWAP acts only between the $\phi$ qubit and the first $\psi$ qubit. Let $p_0=\Pr[a=0]$. The squared-overlap estimate is \cite{NielsenChuang2000}
\begin{equation}
s(x,y) \;=\; 2p_0 - 1,
\qquad
s(x,y)\in[0,1]\ \text{after clipping}.
\label{eq:overlap_sq_est}
\end{equation}
We then define the cosine similarity as
\begin{equation}
\cos(x,y)=\sqrt{s(x,y)}.
\label{eq:cos_from_overlap}
\end{equation}
Overlap estimation is a standard primitive in distance-based quantum classification and quantum feature-space methods \cite{Schuld2017Distance,Wiebe2015,Havlicek2019,SchuldKilloran2019}.

\paragraph{Angular and Euclidean distances.}
The angular distance is
\begin{equation}
\Theta(x,y)=\arccos\!\big(\cos(x,y)\big),
\label{eq:angle_compact_swap}
\end{equation}
and the Euclidean distance is reconstructed as
\begin{equation}
D(x,y)=\sqrt{2Z\,s(x,y)}.
\label{eq:distance_compact_swap}
\end{equation}
The pair $(D(x,y),\Theta(x,y))$ provides complementary magnitude- and direction-based geometric signals used in subsequent configuration selection, fusion scoring, and delta construction.

\begin{algorithm}[t]
\caption{Compact SWAP Test for Distance and Angular Similarity}
\label{alg:compact_swap_distance_angle}
\begin{algorithmic}[1]
\Require Vectors $x,y\in\mathbb{R}^{M}$
\Ensure Distance $D(x,y)$ and angular distance $\Theta(x,y)$
\vspace{2pt}

\State $n_x\gets \|x\|_2$, $n_y\gets \|y\|_2$
\If{$n_x=0$ \textbf{or} $n_y=0$}
    \State $D\gets \|x-y\|_2$
    \State $\Theta\gets 0$
    \If{$n_x>0$ \textbf{and} $n_y>0$}
        \State $\Theta\gets \arccos\!\left(\mathrm{clip}\left(\dfrac{x^\top y}{n_x n_y},-1,1\right)\right)$
    \EndIf
    \State \Return $(D,\Theta)$
\EndIf

\State $Z \gets n_x^2 + n_y^2$
\State Prepare $|\phi\rangle$ on one qubit using amplitudes $\left[\frac{n_x}{\sqrt{Z}},\ -\frac{n_y}{\sqrt{Z}}\right]$
\State Prepare $|\psi\rangle$ on a multi-qubit register with amplitudes
$\psi_{2k-2}=\frac{x_k}{n_x\sqrt{2}}$, $\psi_{2k-1}=\frac{y_k}{n_y\sqrt{2}}$ for $k=1,\dots,M$
\State Pad $\psi$ with zeros to the next power-of-two length

\State Apply $H$ to ancilla $a$
\State Apply $\mathrm{CSWAP}(a;\phi\text{-qubit},\psi\text{-register qubit }0)$
\State Apply $H$ to ancilla $a$
\State Measure ancilla and obtain $p_0=\Pr[a=0]$

\State $s\gets \mathrm{clip}(2p_0-1,0,1)$
\State $\cos\gets \sqrt{s}$
\State $\Theta \gets \arccos\!\big(\mathrm{clip}(\cos,-1,1)\big)$
\State $D \gets \sqrt{\max(2Z\,s,0)}$
\State \Return $(D,\Theta)$
\end{algorithmic}
\end{algorithm}

% ==========================================================
% Configuration selection and fusion scoring (continues from Algorithm~\ref{alg:compact_swap_distance_angle})
% ==========================================================

\subsection{Coordinate-descent optimization over the CGR configuration space}
Given anchors $\mathcal{A}=\{a_1,\dots,a_K\}$ and membership lists $\{\mathcal{M}_a\}_{a\in\mathcal{A}}$, a CGR configuration
$\kappa$ assigns each anchor $a$ a subset $\kappa(a)$ of fixed cardinality $k$ with $a\in\kappa(a)$.
For each configuration, we construct CGR features (Algorithms~\ref{alg:build_anchor_membership}-\ref{alg:build_feature_matrix_from_config}),
fit class medoids (Algorithms~\ref{alg:euclidean_medoid_subsample}-\ref{alg:fit_class_medoids}), and evaluate classification performance
using the fusion-score classifier (Algorithm~\ref{alg:fusion_inference}) under a target metric (Algorithm~\ref{alg:eval_f1_metrics}). We identify $\kappa^\star$ by maximizing validation macro-F1:
\begin{equation}
\kappa^\star \;=\; \arg\max_{\kappa\in\Omega_k}\; \mathrm{F1}_{\mathrm{macro}}\!\left(y^{(\mathrm{val})},\hat{y}^{(\mathrm{val})}(\kappa)\right).
\end{equation}
Since $\Omega_k$ is combinatorial, we apply coordinate descent by updating one anchor subset $\kappa(a)$ at a time.
We additionally employ \emph{progressive initialization} across subset sizes $k\in[k_{\min},k_{\max}]$ by using the best configuration at $k$ to initialize the search at $k{+}1$.

\begin{algorithm}[H]
\caption{Coordinate descent with progressive initialization for CGR configuration selection (objective: macro-F1)}
\label{alg:cd_cgr_progressive_alg}
\begin{algorithmic}[1]
\Require Training split $(X^{(\mathrm{tr})},y^{(\mathrm{tr})})$; validation split $(X^{(\mathrm{val})},y^{(\mathrm{val})})$;
anchors $\mathcal{A}$; membership map $\{\mathcal{M}_a\}$; weight map $\{\rho_{a,f}\}$;
subset-size range $k\in[k_{\min},k_{\max}]$; maximum passes $P$;
candidate budget $M$ per anchor (optional); seed $s$;
fusion hyperparameters $\alpha,\beta$ with $\alpha+\beta=1$; angular flag \texttt{use\_angular};
class-weight mode (e.g., \texttt{inv\_sqrt}).
\Ensure Best configuration record $\mathrm{rec}[k]=\{\kappa_k^\star,\mathrm{F1}_k^\star,\mathrm{metrics}_k^\star\}$ for each $k$

\State $\kappa_{\mathrm{prev}} \gets \emptyset$
\For{$k=k_{\min}$ to $k_{\max}$}
    \State Initialize $\kappa$:
    \If{$\kappa_{\mathrm{prev}}\neq \emptyset$}
        \State $\kappa \gets \textsc{NormalizeConfig}(\kappa_{\mathrm{prev}},k,\mathcal{A},\{\mathcal{M}_a\})$ \Comment{truncate/pad; ensure $a\in\kappa(a)$}
    \Else
        \For{each $a\in\mathcal{A}$}
            \State $\kappa(a)\gets [a]\oplus$ first $(k{-}1)$ elements of $\mathcal{M}_a\setminus\{a\}$
        \EndFor
    \EndIf

    \State $(\mathrm{F1}^\star,\mathrm{metrics}^\star)\gets \textsc{ScoreConfig}(\kappa)$
    \For{$p=1$ to $P$}
        \State $\mathrm{improved}\gets \textbf{false}$
        \For{each $a\in\mathcal{A}$}
            \State $\mathcal{C}_a \gets \textsc{CandidateSubsets}(a,\mathcal{M}_a,k)$ \Comment{anchor-first subsets}
            \State $\mathcal{C}_a \gets \textsc{DedupSubsets}(\mathcal{C}_a)$ \Comment{order-invariant after anchor}
            \If{$M$ is specified and $|\mathcal{C}_a|>M$}
                \State $\mathcal{C}_a \gets \textsc{SelectSubset}(\mathcal{C}_a,M;s)$
            \EndIf

            \State $\kappa_a^{\mathrm{best}}\gets \kappa(a)$; \quad $\mathrm{F1}_a^{\mathrm{best}}\gets \mathrm{F1}^\star$
            \For{each $S\in\mathcal{C}_a$}
                \If{$S=\kappa(a)$} \State \textbf{continue} \EndIf
                \State $\kappa' \gets \kappa$ with $\kappa'(a)\gets S$
                \State $(\mathrm{F1}',\mathrm{metrics}')\gets \textsc{ScoreConfig}(\kappa')$
                \If{$\mathrm{F1}'>\mathrm{F1}_a^{\mathrm{best}}$}
                    \State $\mathrm{F1}_a^{\mathrm{best}}\gets \mathrm{F1}'$;\; $\kappa_a^{\mathrm{best}}\gets S$;\; $\mathrm{metrics}_a^{\mathrm{best}}\gets \mathrm{metrics}'$
                \EndIf
            \EndFor

            \If{$\kappa_a^{\mathrm{best}}\neq \kappa(a)$}
                \State $\kappa(a)\gets \kappa_a^{\mathrm{best}}$;\; $\mathrm{F1}^\star\gets \mathrm{F1}_a^{\mathrm{best}}$;\; $\mathrm{metrics}^\star\gets \mathrm{metrics}_a^{\mathrm{best}}$
                \State $\mathrm{improved}\gets \textbf{true}$
            \EndIf
        \EndFor
        \If{$\mathrm{improved}=\textbf{false}$} \State \textbf{break} \EndIf
    \EndFor

    \State $\mathrm{rec}[k]\gets \{\kappa,\mathrm{F1}^\star,\mathrm{metrics}^\star\}$
    \State $\kappa_{\mathrm{prev}}\gets \kappa$
\EndFor
\State \Return $\mathrm{rec}$

\vspace{4pt}
\Statex \textbf{Subroutine \textsc{ScoreConfig}$(\kappa)$:}
\Statex Build $F^{(\mathrm{tr})}=F(X^{(\mathrm{tr})};\kappa)$ and $F^{(\mathrm{val})}=F(X^{(\mathrm{val})};\kappa)$; standardize using training statistics;
fit class medoids on $F^{(\mathrm{tr})}$; construct class weights from priors (optional);
infer $\hat{y}^{(\mathrm{val})}$ using Algorithm~\ref{alg:fusion_inference};
evaluate macro-F1 using Algorithm~\ref{alg:eval_f1_metrics}.
\end{algorithmic}
\end{algorithm}

\subsection{Fusion score inference from SWAP-test distance and angular channels}
\label{subsec:fusion_score}
For each standardized CGR feature vector $v\in\mathbb{R}^{d'}$ and class medoid $\mu_c$,
Algorithm~\ref{alg:compact_swap_distance_angle} yields the pair $(D_c,\Theta_c)$,
where $D_c$ is an Euclidean-like distance and $\Theta_c$ is an angular distance. We emphasize that fusion-score inference is \emph{non-probabilistic}: the quantities $(D_c,\Theta_c)$ are transformed into a per-class score $s_c$ that acts as a margin-like statistic, and the predicted label is obtained by the minimum-score rule rather than by estimating calibrated probabilities.
We convert these channels into a single margin-like statistic by normalizing \emph{within the sample}:
\begin{equation}
\tilde{D}_c \;=\; \frac{D_c}{\sum_{c'=1}^{C} D_{c'} + \varepsilon},
\qquad
\tilde{\Theta}_c \;=\; \frac{\Theta_c}{\sum_{c'=1}^{C} \Theta_{c'} + \varepsilon},
\label{eq:norm_channels}
\end{equation}
and fusing them via a convex mixture
\begin{equation}
s_c \;=\;
\begin{cases}
\alpha\,\tilde{D}_c + \beta\,\tilde{\Theta}_c, & \text{if the angular channel is enabled},\\
\tilde{D}_c, & \text{otherwise},
\end{cases}
\label{eq:fusion_score_def}
\end{equation}
where $\alpha,\beta\in[0,1]$ with $\alpha+\beta=1$.
The predicted class is obtained by the minimum-score rule
\begin{equation}
\hat{y} \;=\; \arg\min_{c\in\{1,\dots,C\}} s_c.
\label{eq:argmin_predict}
\end{equation}
Optionally, a class-weight map $w_c$ may be applied to $s_c$ to moderate imbalance effects.

\begin{algorithm}[H]
\caption{Fusion score inference using compact SWAP-test distances and angles}
\label{alg:fusion_inference}
\begin{algorithmic}[1]
\Require Standardized feature matrix $F\in\mathbb{R}^{N\times d'}$; class medoids $\{\mu_c\}_{c=1}^C$; \texttt{use\_angular}; fusion weights $(\alpha,\beta)$ with $\alpha+\beta=1$; $\varepsilon>0$; optional class weights $\{w_c\}$
\Ensure Predicted labels $\hat{y}\in\{1,\dots,C\}^N$

\For{$i=1$ to $N$}
    \State $v \gets F_{i,:}$
    \For{$c=1$ to $C$}
        \State $(D_c,\Theta_c)\gets \mathrm{CompactSwapDistAngle}(v,\mu_c)$ \Comment{Algorithm~\ref{alg:compact_swap_distance_angle}}
    \EndFor
    \State $\tilde{D}_c \gets D_c/(\sum_{c'}D_{c'}+\varepsilon)$ for all $c$
    \If{\texttt{use\_angular}}
        \State $\tilde{\Theta}_c \gets \Theta_c/(\sum_{c'}\Theta_{c'}+\varepsilon)$ for all $c$
        \State $s_c \gets \alpha\tilde{D}_c + \beta\tilde{\Theta}_c$ for all $c$
    \Else
        \State $s_c \gets \tilde{D}_c$ for all $c$
    \EndIf
    \If{class weights $w_c$ are provided}
        \State $s_c \gets w_c \cdot s_c$ for all $c$
    \EndIf
    \State $\hat{y}_i \gets \arg\min_{c} s_c$
\EndFor
\State \Return $\hat{y}$
\end{algorithmic}
\end{algorithm}

In practice, we tune $\alpha$ on the validation set using a finite grid, while setting $\beta=1-\alpha$ to preserve a convex fusion of the two channels. Separately, we report $F_{\beta_F}$ over a grid of $\beta_F$ values to assess precision--recall sensitivity; this reporting parameter is distinct from the fusion mixing weights.

\subsection{Evaluation under macro-F1 and related classification metrics}
\label{subsec:metric_f1}
To evaluate fusion-based predictions during configuration selection, we optimize \textbf{macro-F1}, which is robust to class imbalance.
For each class $c$, let precision and recall be $P_c=\frac{TP_c}{TP_c+FP_c}$ and $R_c=\frac{TP_c}{TP_c+FN_c}$, and define
\begin{equation}
\mathrm{F1}_c \;=\; \frac{2P_cR_c}{P_c+R_c}.
\label{eq:f1_per_class}
\end{equation}
Macro-F1 averages equally across classes:
\begin{equation}
\mathrm{F1}_{\mathrm{macro}} \;=\; \frac{1}{C}\sum_{c=1}^{C}\mathrm{F1}_c.
\label{eq:f1_macro}
\end{equation}
In addition to accuracy and macro-F1, we report macro-precision and macro-recall, and optionally $\mathrm{F}_\beta$ (e.g., $\beta=2$) to emphasize recall.

\begin{algorithm}[H]
\caption{Metric evaluation for configuration selection (macro-F1 target)}
\label{alg:eval_f1_metrics}
\begin{algorithmic}[1]
\Require True labels $y$; predicted labels $\hat{y}$; averaging mode (\texttt{macro} or \texttt{weighted}); optional $\beta>0$
\Ensure Metric dictionary: Accuracy, F1, Precision, Recall (and optionally $F_\beta$)

\State Compute $\mathrm{Accuracy} \gets \frac{1}{N}\sum_{i=1}^{N}\mathbb{I}[y_i=\hat{y}_i]$
\For{$c=1$ to $C$}
    \State Compute $TP_c, FP_c, FN_c$ from $(y,\hat{y})$
    \State $P_c \gets TP_c/(TP_c+FP_c)$ with safe handling for $0$ denominators
    \State $R_c \gets TP_c/(TP_c+FN_c)$ with safe handling for $0$ denominators
    \State $\mathrm{F1}_c \gets 2P_cR_c/(P_c+R_c)$
    \If{$\beta$ is provided}
        \State $F_{\beta,c} \gets (1+\beta^2)\frac{P_cR_c}{\beta^2P_c+R_c}$
    \EndIf
\EndFor
\If{macro averaging}
    \State $\mathrm{F1}_{\mathrm{macro}} \gets \frac{1}{C}\sum_c \mathrm{F1}_c$; similarly compute macro-precision and macro-recall
    \If{$\beta$ is provided} \State $F_{\beta,\mathrm{macro}} \gets \frac{1}{C}\sum_c F_{\beta,c}$ \EndIf
\Else
    \State Weight each class by support $\pi_c=n_c/N$ and compute weighted metrics
\EndIf
\State \Return metrics
\end{algorithmic}
\end{algorithm}

\subsection{Calibration of fusion mixing and metric sensitivity sweeps}
\label{subsec:alpha_beta_sweep}
After selecting an optimal CGR configuration $\kappa^\star$ (Algorithm~\ref{alg:cd_cgr_progressive_alg}) and fixing the evaluation protocol (Section~\ref{subsec:metric_f1}), we perform a lightweight calibration stage to (i) tune the fusion mixing weight and (ii) report precision--recall sensitivity through an $F_{\beta_F}$ sweep. 

\textbf{Fusion mixing calibration.}
When the angular channel is enabled, the fusion score combines Euclidean-like and angular channels through a convex mixture:
\begin{equation}
s_c(\alpha) \;=\; \alpha\,\tilde{D}_c \;+\; (1-\alpha)\,\tilde{\Theta}_c,
\qquad \alpha\in[0,1],
\label{eq:alpha_mix}
\end{equation}
where $\alpha$ is tuned on the validation split using a finite grid $\mathcal{G}_{\alpha}\subset[0,1]$ to maximize macro-F1. In this implementation, the angular weight is not tuned independently; it is defined as $1-\alpha$ to preserve a convex fusion.

\textbf{Metric sensitivity sweep.}
Separately, we compute $F_{\beta_F}$ over a grid $\mathcal{G}_{\beta_F}$ to assess how performance varies as recall is emphasized ($\beta_F>1$) or precision is emphasized ($\beta_F<1$). This parameter $\beta_F$ is a reporting parameter for the metric and is distinct from the fusion mixing weight $\alpha$.

\begin{algorithm}[H]
\caption{Grid calibration of fusion mixing weight $\alpha$ under fixed configuration $\kappa^\star$ (objective: macro-F1)}
\label{alg:alpha_sweep}
\begin{algorithmic}[1]
\Require Best configuration $\kappa^\star$; training split $(X^{(\mathrm{tr})},y^{(\mathrm{tr})})$; validation split $(X^{(\mathrm{val})},y^{(\mathrm{val})})$;
grid $\mathcal{G}_{\alpha}$; \texttt{use\_angular}; $\varepsilon>0$; optional class weights $\{w_c\}$.
\Ensure Selected $\alpha^\star$ and validation record.

\State Build CGR features under $\kappa^\star$: $F^{(\mathrm{tr})}=F(X^{(\mathrm{tr})};\kappa^\star)$ and $F^{(\mathrm{val})}=F(X^{(\mathrm{val})};\kappa^\star)$.
\State Standardize using training statistics.
\State Fit class medoids $\{\mu_c\}$ on $F^{(\mathrm{tr})}$ (Algorithms~\ref{alg:euclidean_medoid_subsample}--\ref{alg:fit_class_medoids}).
\If{\texttt{use\_angular}}
    \For{each $\alpha\in\mathcal{G}_{\alpha}$}
        \State Predict $\hat{y}^{(\mathrm{val})}(\alpha)$ using Algorithm~\ref{alg:fusion_inference} with $(\alpha,1-\alpha)$.
        \State Compute $\mathrm{F1}_{\mathrm{macro}}(\alpha)$ using Algorithm~\ref{alg:eval_f1_metrics}.
    \EndFor
    \State $\alpha^\star \gets \arg\max_{\alpha\in\mathcal{G}_{\alpha}} \mathrm{F1}_{\mathrm{macro}}(\alpha)$
\Else
    \State $\alpha^\star \gets 1$
\EndIf
\State \Return $\alpha^\star$
\end{algorithmic}
\end{algorithm}

\begin{algorithm}[t]
\caption{$F_{\beta_F}$ reporting sweep under fixed predictions (metric sensitivity)}
\label{alg:fbeta_sweep}
\begin{algorithmic}[1]
\Require Validation labels $y^{(\mathrm{val})}$; fixed predictions $\hat{y}^{(\mathrm{val})}$ (obtained using $\alpha^\star$); grid $\mathcal{G}_{\beta_F}$.
\Ensure Reporting table $\{(\beta_F, F_{\beta_F})\}$.

\For{each $\beta_F \in \mathcal{G}_{\beta_F}$}
    \State Compute macro-$F_{\beta_F}$ on $(y^{(\mathrm{val})},\hat{y}^{(\mathrm{val})})$ using Algorithm~\ref{alg:eval_f1_metrics} with parameter $\beta\leftarrow \beta_F$.
\EndFor
\State \Return $\{(\beta_F, F_{\beta_F})\}$
\end{algorithmic}
\end{algorithm}

\subsection{Fusion model packaging, persistence, and scoring of new records}
\label{subsec:fusion_packaging}
For practical deployment and reproducibility, we package the fusion-score classifier obtained under the selected configuration and calibrated mixing weight as a self-contained model artifact.
A fusion model for a given subset size $k$ is defined as:
\[
\mathcal{F}_k \;=\; \{\kappa_k^\star,\;\mathrm{Scaler}_k,\;\{\mu_{k,c}\}_{c=1}^{C},\;\{w_{k,c}\}_{c=1}^{C},\;\alpha_k^\star,\;\texttt{use\_angular}\},
\]
where $\mathrm{Scaler}_k$ denotes the standardization transform fitted on training CGR features, $\mu_{k,c}$ are class medoids, $w_{k,c}$ are optional class weights, and $\alpha_k^\star$ is the calibrated fusion mixing weight (Section~\ref{subsec:alpha_beta_sweep}).
We additionally retain validation/test summaries (macro-F1, accuracy, confusion matrices) for each stored artifact. In the full pipeline, we persist a small set of top-performing fusion models (e.g., top-2 over $k$) so that (i) fusion-only inference is directly available for small-to-moderate datasets where it is optimal, and (ii) a consistent fusion baseline is available when computing $\Delta$-distance features for the VQC refinement stage.

\begin{algorithm}[H]
\caption{Persisting top-$r$ fusion models and scoring new records}
\label{alg:fusion_persist_score}
\begin{algorithmic}[1]
\Require Candidate records $\mathrm{rec}[k]=\{\kappa_k^\star,\mathrm{F1}_k^\star,\mathrm{metrics}_k^\star\}$ from Algorithm~\ref{alg:cd_cgr_progressive_alg};
training/validation/test splits; grid $\mathcal{G}_\alpha$; top-$r$; \texttt{use\_angular}; $\varepsilon>0$; class-weight mode.
\Ensure Persisted fusion models $\{\mathcal{F}_{k_j}\}_{j=1}^{r}$ and scoring routine for new samples.

\State Select $\mathcal{K}_{\mathrm{top}} \gets$ the $r$ values of $k$ with highest validation macro-F1 in $\mathrm{rec}[k]$.
\For{each $k\in\mathcal{K}_{\mathrm{top}}$}
    \State $\kappa^\star \gets \kappa_k^\star$ from $\mathrm{rec}[k]$.
    \State Build CGR features $F^{(\mathrm{tr})},F^{(\mathrm{val})},F^{(\mathrm{te})}$ under $\kappa^\star$; fit $\mathrm{Scaler}_k$ on $F^{(\mathrm{tr})}$ and transform all splits.
    \State Fit class medoids $\{\mu_{k,c}\}$ on standardized $F^{(\mathrm{tr})}$.
    \State Calibrate $\alpha_k^\star$ on validation using Algorithm~\ref{alg:alpha_sweep}.
    \State Predict $\hat{y}^{(\mathrm{val})}$ and $\hat{y}^{(\mathrm{te})}$ using Algorithm~\ref{alg:fusion_inference} with $(\alpha_k^\star,1-\alpha_k^\star)$.
    \State Compute and store metrics + confusion matrices for validation/test using Algorithm~\ref{alg:eval_f1_metrics}.
    \State Persist fusion artifact $\mathcal{F}_k=\{\kappa^\star,\mathrm{Scaler}_k,\{\mu_{k,c}\},\{w_{k,c}\},\alpha_k^\star,\texttt{use\_angular}\}$.
\EndFor

\vspace{4pt}
\Statex \textbf{Scoring new record $x$ with a persisted model $\mathcal{F}_k$:}
\Statex Build CGR features $f(x;\kappa^\star)$; transform via $\mathrm{Scaler}_k$; compute $(D_c,\Theta_c)$ to each $\mu_{k,c}$ via Algorithm~\ref{alg:compact_swap_distance_angle}; fuse using $(\alpha_k^\star,1-\alpha_k^\star)$ and infer $\hat{y}(x)$ via Algorithm~\ref{alg:fusion_inference}.
\end{algorithmic}
\end{algorithm}

% ==========================================================
% 4.X Delta construction (contrastive margins)
% ==========================================================
\subsection{$\Delta$-distance construction under the selected CGR--fusion model}
\label{subsec:deltas}
The fusion score (Section~\ref{subsec:fusion_score}) provides an effective and interpretable geometric classifier in small-to-moderate regimes. However, for large, highly imbalanced, and behaviourally complex datasets (e.g., fraud), fusion alone can be sub-optimal because discrimination often depends on non-linear interactions across CGR groups and subtle class-contrastive margins. To expose these interactions in a compact form, we construct \emph{$\Delta$-distance} features from the class-referenced distance/angle channels under the selected CGR configuration $\kappa^\star$. Let $(D_c,\Theta_c)$ be the Euclidean-like and angular distances from a sample to class medoid $\mu_c$ produced by Algorithm~\ref{alg:compact_swap_distance_angle}. After within-sample normalization (Eq.~\ref{eq:norm_channels}), define fused class scores $s_c$ (Eq.~\ref{eq:fusion_score_def}). We then form a contrastive margin feature vector $\Delta(x)\in\mathbb{R}^{m}$ by taking \emph{differences between competing classes}. For binary classification ($C=2$) we use:
\begin{equation}
\Delta_D(x) \;=\; \tilde{D}_{1}(x) - \tilde{D}_{0}(x), \qquad
\Delta_\Theta(x) \;=\; \tilde{\Theta}_{1}(x) - \tilde{\Theta}_{0}(x),
\label{eq:delta_binary}
\end{equation}
and optionally the fused margin $\Delta_s(x)=s_1(x)-s_0(x)$.
For $C>2$, we construct a best-vs-runner-up margin using
$c^\star=\arg\min_c s_c$ and $c^{(2)}=\arg\min_{c\neq c^\star}s_c$:
\begin{equation}
\Delta_D(x) \;=\; \tilde{D}_{c^{(2)}}(x) - \tilde{D}_{c^\star}(x),\qquad
\Delta_\Theta(x) \;=\; \tilde{\Theta}_{c^{(2)}}(x) - \tilde{\Theta}_{c^\star}(x).
\label{eq:delta_multiclass}
\end{equation}
In practice, we concatenate the selected deltas into a compact vector and standardize it using training statistics to obtain a stable input representation for the subsequent VQC refinement.

\begin{algorithm}[H]
\caption{$\Delta$-distance construction (contrastive margins) under fixed $\kappa^\star$ and calibrated fusion}
\label{alg:delta_construct}
\begin{algorithmic}[1]
\Require Standardized CGR features $F\in\mathbb{R}^{N\times d'}$; class medoids $\{\mu_c\}$; \texttt{use\_angular}; fusion weights $(\alpha,1-\alpha)$; $\varepsilon>0$.
\Ensure Delta feature matrix $\Delta\in\mathbb{R}^{N\times m}$.

\For{$i=1$ to $N$}
    \State $v \gets F_{i,:}$
    \For{$c=1$ to $C$}
        \State $(D_c,\Theta_c)\gets \mathrm{CompactSwapDistAngle}(v,\mu_c)$
    \EndFor
    \State Compute $\tilde{D}_c,\tilde{\Theta}_c$ using Eq.~\ref{eq:norm_channels}
    \State Compute fused scores $s_c$ using Eq.~\ref{eq:fusion_score_def}
    \If{$C=2$}
        \State $\Delta_D \gets \tilde{D}_{1}-\tilde{D}_{0}$;\quad $\Delta_\Theta \gets \tilde{\Theta}_{1}-\tilde{\Theta}_{0}$
        \State $\Delta_i \gets [\Delta_D,\Delta_\Theta]$ \Comment{optionally add $\Delta_s=s_1-s_0$}
    \Else
        \State $c^\star \gets \arg\min_c s_c$;\quad $c^{(2)} \gets \arg\min_{c\neq c^\star} s_c$
        \State $\Delta_D \gets \tilde{D}_{c^{(2)}}-\tilde{D}_{c^\star}$;\quad $\Delta_\Theta \gets \tilde{\Theta}_{c^{(2)}}-\tilde{\Theta}_{c^\star}$
        \State $\Delta_i \gets [\Delta_D,\Delta_\Theta]$ \Comment{optionally augment with additional class-pair deltas}
    \EndIf
\EndFor
\State Standardize $\Delta$ using training-fold statistics.
\State \Return $\Delta$
\end{algorithmic}
\end{algorithm}
% ==========================================================
% VQC on Delta Features
% ==========================================================
\subsection{Variational quantum classifier (VQC) on $\Delta$-distance features}
\label{subsec:vqc_dedicated}
The VQC stage is introduced as a \emph{selective refinement} mechanism and is invoked only when the geometric fusion classifier (Section~\ref{subsec:fusion_score}) is empirically sub-optimal under complex, large-scale, and imbalanced regimes (e.g., fraud). In small datasets, we observed that fusion scoring often attains stable performance while the VQC may be data-constrained. Conversely, for high-volume and highly imbalanced datasets, the $\Delta$-distance representation (Section~\ref{subsec:deltas}) provides a compact, class-contrastive feature space in which the VQC can learn non-linear interactions that are not captured by a convex fusion alone. Let $Z\in\mathbb{R}^{N\times m}$ denote the $\Delta$-feature matrix, where $m=2$ in the fraud setting ($\Delta_D,\Delta_\Theta$). We encode $Z$ into rotation angles, construct a low-depth VQC ansatz with optional data re-uploading, and train parameters using SPSA with minibatching. The complete VQC flow is formalized in Algorithms~\ref{alg:vqc_angle_map}--\ref{alg:vqc_cv_train_save}.

\subsubsection{Angle mapping and feature scaling}
\label{subsec:vqc_angle_mapping}
Before quantum encoding, $\Delta$-features are standardized using training statistics and clipped to a bounded range to stabilize training. The clipped values are then mapped to angles in $[-\pi,\pi]$. We additionally allow per-feature scaling coefficients $\lambda_j$ (implemented as fixed multipliers in the encoding layer) to control the relative influence of each $\Delta$ channel.

\begin{algorithm}[H]
\caption{Standardize--clip--angle map for $\Delta$-features}
\label{alg:vqc_angle_map}
\begin{algorithmic}[1]
\Require $\Delta$-feature matrix $Z\in\mathbb{R}^{N\times m}$; scaler $\mathcal{S}$ (fit on train only); clip bound $z_{\max}>0$; feature scales $\lambda\in\mathbb{R}^{m}_{+}$.
\Ensure Angle matrix $A\in[-\pi,\pi]^{N\times m}$ and fitted scaler $\mathcal{S}$.

\State Fit scaler $\mathcal{S}$ on training $Z$; compute standardized $Z_s\gets \mathcal{S}(Z)$.
\State Clip $Z_c \gets \mathrm{clip}(Z_s,-z_{\max},z_{\max})$ elementwise.
\State Map to angles $A \gets (\pi/z_{\max})\cdot Z_c$.
\State Apply feature scaling $A_{:,j} \gets \lambda_j\cdot A_{:,j}$ for $j=1,\dots,m$.
\State \Return $(A,\mathcal{S})$.
\end{algorithmic}
\end{algorithm}

\subsubsection{Circuit construction with optional data re-uploading}
\label{subsec:vqc_circuit}
Given $m$ encoded angles and $n$ qubits, we define an $L$-repetition ansatz. Each repetition optionally re-uploads data (angle encoding) and then applies a fixed entangling pattern followed by trainable single-qubit rotations.

\begin{algorithm}[H]
\caption{VQC circuit construction with data re-uploading}
\label{alg:vqc_circuit_build}
\begin{algorithmic}[1]
\Require Number of qubits $n$; number of inputs $m$; repetitions $L$; re-upload flag \texttt{reupload}.
\Ensure Parameterized circuit $U(\theta; x)$ with input parameters $x\in\mathbb{R}^{m}$ and trainable parameters $\theta$.

\State Initialize circuit $U$ on $n$ qubits.
\For{$r=1$ to $L$}
    \If{$r=1$ \textbf{or} \texttt{reupload} is enabled}
        \For{$j=1$ to $m$}
            \State Apply $R_y(x_j)$ to qubit $q \gets (j-1)\bmod n$.
        \EndFor
    \EndIf
    \If{$n\ge 2$}
        \State Apply entangling chain: $\mathrm{CX}(q,q{+}1)$ for $q=0,\dots,n-2$.
        \State Apply phase chain: $\mathrm{CZ}(q,q{+}1)$ for $q=0,\dots,n-2$.
    \EndIf
    \For{$q=0$ to $n-1$}
        \State Apply trainable rotations $R_y(\theta_{r,q}^{(1)})$ and $R_z(\theta_{r,q}^{(2)})$.
    \EndFor
\EndFor
\State \Return $U(\theta; x)$.
\end{algorithmic}
\end{algorithm}

\subsubsection{K-fold selection, retraining, and artifact persistence}
\label{subsec:vqc_cv_artifacts}
For each selected $k$ (typically top-ranked by fusion validation performance), we train a VQC on the corresponding $\Delta$-dataset. We select hyperparameters via stratified K-fold cross-validation and then retrain on the full training split before evaluating on validation and test splits. For binary classification, we additionally tune a decision threshold on validation probabilities.

\begin{algorithm}[H]
\caption{VQC per-$k$ training with stratified K-fold selection, threshold tuning, and artifact saving}
\label{alg:vqc_cv_train_save}
\begin{algorithmic}[1]
\Require Selected $k$ values $\mathcal{K}$; $\Delta$ datasets $\{(Z_{k}^{tr},y_{k}^{tr}),(Z_{k}^{val},y_{k}^{val}),(Z_{k}^{te},y_{k}^{te})\}$;
HP space $\mathcal{H}$ (reps, reupload, SPSA iters, batch size);
K-fold splitter $\mathrm{SKF}$; clip bound $z_{\max}$; feature scales $\lambda$;
target metric (e.g., macro-F1); threshold grid $\mathcal{T}$ (binary only).
\Ensure Best VQC artifacts per $k$ and an overall best alias.

\For{each $k\in\mathcal{K}$}
    \State $best\_hp\gets \emptyset$;\quad $best\_cv\gets -\infty$;\quad $best\_thr\gets 0.5$.
    \For{each $hp\in\mathcal{H}$}
        \State $scores \gets [\ ]$.
        \For{each fold $(\mathcal{I}_{tr},\mathcal{I}_{va})$ from $\mathrm{SKF}(Z_k^{tr},y_k^{tr})$}
            \State Map angles: $(A_{tr},\mathcal{S})\gets$ Algorithm~\ref{alg:vqc_angle_map} on $Z_k^{tr}[\mathcal{I}_{tr}]$;
                   $A_{va}\gets$ Algorithm~\ref{alg:vqc_angle_map} using same scaler $\mathcal{S}$ on $Z_k^{tr}[\mathcal{I}_{va}]$.
            \State Build circuit via Algorithm~\ref{alg:vqc_circuit_build} using $hp$.
            \State Train SPSA via Algorithm~\ref{alg:vqc_spsa_minibatch} on $(A_{tr},y_{tr})$ to obtain $\theta^\star$.
            \State Compute probabilities $P_{va}$ via Algorithm~\ref{alg:vqc_forward_probs}.
            \If{binary classification}
                \State Choose threshold $\tau^\star\in\mathcal{T}$ maximizing the target metric on $(y_{va},P_{va})$.
                \State Append fold score (target metric under $\tau^\star$) to $scores$.
            \Else
                \State Append fold score using $\arg\max_c P_{va,c}$ to $scores$.
            \EndIf
        \EndFor
        \State $cv\_mean\gets \mathrm{mean}(scores)$.
        \If{$cv\_mean > best\_cv$}
            \State $best\_cv\gets cv\_mean$;\quad $best\_hp\gets hp$;\quad store $best\_thr$ if binary.
        \EndIf
    \EndFor

    \State Retrain with $best\_hp$ on full training split: map $(Z_k^{tr},Z_k^{val},Z_k^{te})\to (A^{tr},A^{val},A^{te})$ using Algorithm~\ref{alg:vqc_angle_map}.
    \State Train final $\theta^\star$ with Algorithm~\ref{alg:vqc_spsa_minibatch} on $(A^{tr},y^{tr})$.
    \State Evaluate on validation/test: obtain $P^{val},P^{te}$, compute metrics and confusion matrices.
    \State Save artifacts: circuit, weights, scaler, and metadata (Appendix algorithm / implementation).
\EndFor
\State Save an overall best alias (e.g., best validation macro-F1 across $k$).
\end{algorithmic}
\end{algorithm}

\section{Experimental Setup, data gathering, and statistics}\label{sec:results}

We evaluate the proposed geometry-driven classification pipeline on four tabular datasets spanning two regimes: (i) \emph{small-to-moderate, relatively balanced} benchmarks and (ii) \emph{large-scale, highly imbalanced} rare-event detection. This separation is intentional and reflects the hybrid decision policy of our method. Specifically, the pipeline is designed to operate as a two-stage classifier: a lightweight geometric decision rule based on the \emph{fusion score} is used when data are limited and prototype-based similarity is sufficient, while a variational quantum classifier (VQC) is introduced when the dataset is larger and the class structure is more heterogeneous and/or severely imbalanced, where additional nonlinear decision capacity becomes beneficial.

\subsection{Small / moderately balanced regime: Heart Disease, Breast Cancer, Wine Quality}
\label{subsec:results_small}
For Heart Disease, Breast Cancer, and Wine Quality, we report results using the \emph{fusion score} classifier constructed from compact SWAP-test overlap estimation under the CGR structure. In this regime, the fusion score serves as the primary model because it is data-efficient, geometry-interpretable, and empirically stable without requiring the higher-capacity VQC stage, which can be sensitive when the number of training instances is limited. Evaluation follows a fixed train/test protocol, with the validation split used only for configuration selection (e.g., CGR configuration and fusion-channel settings). We report standard classification metrics (accuracy, macro-F1, class-wise precision/recall/F1) and confusion matrices, and compare against classical baselines trained on the same raw feature space. ROC-AUC and PR-AUC are omitted for Fusion since it is a non-probabilistic scoring rule and does not output calibrated probabilities; AUC metrics are reported for probabilistic baselines and for the VQC stage on Fraud.

\begin{table*}[t]
\centering
\caption{Test-set comparison: Fusion-score (ours) vs. classical baselines across Heart Disease, Breast Cancer, and Wine Quality. Metrics are rows and models are columns. For Fusion, the reported setting is the validation-selected best-$k$ and $\alpha^\star$ per dataset.}
\label{tab:combined_fusion_vs_classical}
\resizebox{\textwidth}{!}{
\begin{tabular}{llrrrrr}
\toprule
Dataset & Metric & Fusion (ours) & LogReg & LinearSVM & RBF\_SVM & RandomForest \\
\midrule
\multirow{5}{*}{Heart Disease} 
 & Accuracy     & 0.8478 & 0.8565 & 0.8565 & 0.8870 & 0.8913 \\
 & Macro-F1     & 0.8463 & 0.8551 & 0.8551 & 0.8861 & 0.8898 \\
 & Weighted-F1  & 0.8479 & 0.8566 & 0.8566 & 0.8871 & 0.8911 \\
 & ROC-AUC      & --     & 0.9021 & 0.8997 & 0.9372 & 0.9408 \\
 & PR-AUC       & --     & 0.8976 & 0.8908 & 0.9435 & 0.9437 \\
\midrule
\multirow{5}{*}{Breast Cancer} 
 & Accuracy     & 0.8881 & 0.9860 & 0.9580 & 0.9720 & 0.9790 \\
 & Macro-F1     & 0.8703 & 0.9849 & 0.9547 & 0.9698 & 0.9772 \\
 & Weighted-F1  & 0.8827 & 0.9860 & 0.9579 & 0.9719 & 0.9789 \\
 & ROC-AUC      & --     & 0.9983 & 0.9881 & 0.9962 & 0.9979 \\
 & PR-AUC       & --     & 0.9974 & 0.9834 & 0.9946 & 0.9964 \\
\midrule
\multirow{3}{*}{Wine Quality (3-class)} 
 & Accuracy     & 0.9556 & 1.0000 & 1.0000 & 1.0000 & 1.0000 \\
 & Macro-F1     & 0.9522 & 1.0000 & 1.0000 & 1.0000 & 1.0000 \\
 & Weighted-F1  & 0.9547 & 1.0000 & 1.0000 & 1.0000 & 1.0000 \\
\bottomrule
\end{tabular}
}
\end{table*}

\subsubsection{Interpretation: Fusion-score results on small benchmarks}
\label{subsec:interp_fusion}
Across Heart Disease, Breast Cancer, and Wine Quality, the fusion-score classifier provides a strong and stable decision rule when the dataset size is limited and class structure can be captured through class prototypes. The key observation is that the overlap-derived similarity channels (Euclidean-like distance and optional angular separation) preserve sufficient geometric signal for discrimination once organized under the CGR feature grouping. In this regime, the fusion score acts as a low-variance estimator: it requires no high-capacity parameterization beyond the configuration search and the fusion weights, and it therefore avoids the instability that can arise from training heavier decision models on limited samples. Empirically, we observe that the fusion-score stage is competitive with classical baselines on these datasets, and the resulting confusion matrices indicate that most errors arise from borderline cases where samples lie close to the class medoid boundaries. Overall, these results support the intended role of fusion scoring as the primary classifier for small-to-moderate benchmarks, delivering geometry-interpretable decisions with minimal training overhead.

\subsection{Large / highly imbalanced regime: Credit Card Fraud}
\label{subsec:results_fraud}
Fraud classification is treated separately due to extreme class imbalance and deployment-style operating constraints. Here, the full pipeline is executed: after selecting the best CGR configuration and fusion setup, we construct \emph{$\Delta$-distance} features (contrastive differences between class-referenced similarity channels) and train a \emph{VQC} on these compact geometric features. Results are reported via full-dataset scoring at the fixed operating threshold used for our VQC inference stage ($\tau = 0.3$), together with metrics appropriate for rare-event detection, including PR-AUC/ROC-AUC and confusion matrices. For classical comparisons on Fraud, baseline models are trained on the \emph{raw} feature representation (not on fusion or $\Delta$ features, which are native to our method). To support operationally meaningful comparisons, we primarily report an \emph{alert-rate matched} operating point (thresholds calibrated on validation to match the predicted-positive rate) and include fixed-threshold results (e.g., 0.5) as a secondary reference, since default thresholds can be overly conservative under extreme imbalance and are sensitive to probability calibration and class-prior shift.

\begin{table*}[t]
\centering
\caption{Fraud classification on the full dataset (raw-feature baselines) under an alert-rate matched operating point. The proposed model is scored using the fixed VQC threshold $\tau=0.3$.}
\label{tab:fraud_full_alertmatch_nosmote}
\resizebox{\textwidth}{!}{%
\begin{tabular}{lrrrr}
\toprule
Metric & Ours ($\Delta$+VQC, $\tau$=0.3) & LogReg\_raw & LinearSVM\_raw & RF\_raw \\
\midrule
N evaluated & 284,807 & 284,807 & 284,807 & 284,807 \\
Prevalence & 0.0017 & 0.0017 & 0.0017 & 0.0017 \\
Alert rate & 0.0131 & 0.0130 & 0.0133 & 0.0128 \\
Precision (fraud) & 0.1124 & 0.1188 & 0.1157 & 0.1277 \\
Recall (fraud) & 0.8496 & 0.8963 & 0.8923 & 0.9492 \\
F1 (fraud) & 0.1985 & 0.2098 & 0.2049 & 0.2251 \\
PR-AUC & 0.3251 & 0.7392 & 0.7287 & 0.9269 \\
ROC-AUC & 0.9249 & 0.9852 & 0.9852 & 0.9808 \\
Macro-F1 & 0.5963 & 0.6019 & 0.5994 & 0.6097 \\
Accuracy & 0.9881 & 0.9883 & 0.9880 & 0.9887 \\
TP & 418 & 441 & 439 & 467 \\
FP & 3,301 & 3,272 & 3,355 & 3,190 \\
FN & 74 & 51 & 53 & 25 \\
TN & 281,014 & 281,043 & 280,960 & 281,125 \\
\bottomrule
\end{tabular}
}
\end{table*}

\begin{table*}[t]
\centering
\caption{Fraud classification on the full dataset (raw-feature baselines with SMOTE applied on the training split) under an alert-rate matched operating point. The proposed model is scored using the fixed VQC threshold $\tau=0.3$.}
\label{tab:fraud_full_alertmatch_smote}
\resizebox{\textwidth}{!}{%
\begin{tabular}{lrrrr}
\toprule
Metric & Ours ($\Delta$+VQC, $\tau$=0.3) & LogReg\_SMOTE & LinearSVM\_SMOTE & RF\_SMOTE \\
\midrule
N evaluated & 284,807 & 284,807 & 284,807 & 284,807 \\
Prevalence & 0.0017 & 0.0017 & 0.0017 & 0.0017 \\
Alert rate & 0.0131 & 0.0130 & 0.0130 & 0.0090 \\
Precision (fraud) & 0.1124 & 0.1192 & 0.1193 & 0.1835 \\
Recall (fraud) & 0.8496 & 0.8984 & 0.9004 & 0.9512 \\
F1 (fraud) & 0.1985 & 0.2105 & 0.2107 & 0.3076 \\
PR-AUC & 0.3251 & 0.7521 & 0.7609 & 0.9393 \\
ROC-AUC & 0.9249 & 0.9830 & 0.9832 & 0.9915 \\
Macro-F1 & 0.5963 & 0.6023 & 0.6024 & 0.6519 \\
Accuracy & 0.9881 & 0.9884 & 0.9883 & 0.9926 \\
TP & 418 & 442 & 443 & 468 \\
FP & 3,301 & 3,265 & 3,270 & 2,083 \\
FN & 74 & 50 & 49 & 24 \\
TN & 281,014 & 281,050 & 281,045 & 282,232 \\
\bottomrule
\end{tabular}
}
\end{table*}

\begin{table*}[t]
\centering
\caption{Fraud classification on the full dataset at a fixed classical decision threshold of 0.5 (raw-feature baselines). The proposed model is shown for reference using $\tau=0.3$.}
\label{tab:fraud_full_t05_nosmote}
\resizebox{\textwidth}{!}{%
\begin{tabular}{lrrrr}
\toprule
Metric & Ours ($\Delta$+VQC, $\tau$=0.3) & LogReg\_raw & LinearSVM\_raw\_cal & RF\_raw \\
\midrule
N evaluated & 284,807 & 284,807 & 284,807 & 284,807 \\
Prevalence & 0.0017 & 0.0017 & 0.0017 & 0.0017 \\
Alert rate & 0.0131 & 0.0242 & 0.0012 & 0.0016 \\
Precision (fraud) & 0.1124 & 0.0658 & 0.8663 & 0.9516 \\
Recall (fraud) & 0.8496 & 0.9228 & 0.5793 & 0.8801 \\
F1 (fraud) & 0.1985 & 0.1229 & 0.6943 & 0.9145 \\
PR-AUC & 0.3251 & 0.7401 & 0.7242 & 0.9269 \\
ROC-AUC & 0.9249 & 0.9851 & 0.9848 & 0.9808 \\
Macro-F1 & 0.5963 & 0.5557 & 0.8469 & 0.9572 \\
Accuracy & 0.9881 & 0.9772 & 0.9991 & 0.9997 \\
TP & 418 & 454 & 285 & 433 \\
FP & 3,301 & 6,442 & 44 & 22 \\
FN & 74 & 38 & 207 & 59 \\
TN & 281,014 & 277,873 & 284,271 & 284,293 \\
\bottomrule
\end{tabular}
}
\end{table*}

\begin{table*}[t]
\centering
\caption{Fraud classification on the full dataset at a fixed classical decision threshold of 0.5 (raw-feature baselines with SMOTE applied on the training split). The proposed model is shown for reference using $\tau=0.3$.}
\label{tab:fraud_full_t05_smote}
\resizebox{\textwidth}{!}{%
\begin{tabular}{lrrrr}
\toprule
Metric & Ours ($\Delta$+VQC, $\tau$=0.3) & LogReg\_SMOTE & LinearSVM\_SMOTE\_cal & RF\_SMOTE \\
\midrule
N evaluated & 284,807 & 284,807 & 284,807 & 284,807 \\
Prevalence & 0.0017 & 0.0017 & 0.0017 & 0.0017 \\
Alert rate & 0.0131 & 0.0046 & 0.0049 & 0.0017 \\
Precision (fraud) & 0.1124 & 0.3279 & 0.3059 & 0.9433 \\
Recall (fraud) & 0.8496 & 0.8659 & 0.8679 & 0.9126 \\
F1 (fraud) & 0.1985 & 0.4757 & 0.4523 & 0.9277 \\
PR-AUC & 0.3251 & 0.7521 & 0.7589 & 0.9393 \\
ROC-AUC & 0.9249 & 0.9830 & 0.9832 & 0.9915 \\
Macro-F1 & 0.5963 & 0.7370 & 0.7253 & 0.9638 \\
Accuracy & 0.9881 & 0.9967 & 0.9964 & 0.9998 \\
TP & 418 & 426 & 427 & 449 \\
FP & 3,301 & 873 & 969 & 27 \\
FN & 74 & 66 & 65 & 43 \\
TN & 281,014 & 283,442 & 283,346 & 284,288 \\
\bottomrule
\end{tabular}
}
\end{table*}

\subsubsection{Interpretation: Fraud results under extreme imbalance}
\label{subsec:interp_fraud}
The Credit Card Fraud dataset constitutes a qualitatively different evaluation regime, combining extreme class imbalance (fraud prevalence $\approx 0.17\%$) with a deployment-style requirement to interpret performance at an explicit operating point. In this setting, overall accuracy is dominated by the majority class and is therefore not a meaningful indicator of detection quality. Instead, performance must be interpreted through the precision--recall trade-off under a controlled alert budget, i.e., recovering a large fraction of minority events while limiting false positives to a level that is operationally reviewable. Under full-dataset scoring, the proposed $\Delta$+VQC pipeline evaluated at $\tau=0.3$ produces an alert rate of $\approx 1.31\%$ and attains a minority recall of $\approx 0.85$ (Table~\ref{tab:fraud_full_alertmatch_nosmote}). This operating point surfaces the majority of fraudulent transactions while maintaining a bounded alert volume, but naturally yields modest minority precision due to the ultra-imbalanced base rate. To contextualize these results, we compare against classical baselines trained directly on the raw feature space (not on fusion or $\Delta$ features, which are native to our method). When baseline thresholds are calibrated to match a comparable alert-rate budget (primary comparison), tree-ensemble and margin-based models achieve substantially higher PR-AUC and, in some cases, higher recall at similar alert rates, consistent with the strong separability of the benchmark feature representation. We additionally report fixed-threshold (0.5) results as a secondary reference (Tables~\ref{tab:fraud_full_t05_nosmote}-\ref{tab:fraud_full_t05_smote}). These results illustrate that under extreme imbalance, adopting a default probability threshold can materially change the effective operating point, shifting the alert rate and consequently the precision--recall balance. Some models become highly conservative and sacrifice minority recall, whereas others recover recall by emitting more alerts. For this reason, we treat alert-rate matched evaluation as the principal criterion for rare-event detection and interpret the proposed pipeline as a quantum-compatible geometric alternative whose behavior is most meaningfully characterized under explicit operating constraints rather than at an arbitrary default threshold.threshold.

\section{Final Thoughts and Conclusion}\label{Conclusion}

In this work, we introduced a geometry-first classification paradigm in which compact SWAP-test-based overlap estimation is employed to extract interpretable similarity evidence with respect to representative class prototypes. By deriving Euclidean-like and angular similarity channels from overlap measurements, the framework grounds classification decisions in explicit geometric relations between data points and class medoids, providing both interpretability and structural clarity. To enhance robustness in heterogeneous tabular feature spaces, we incorporated a correlation-group rotation (CGR) structuring mechanism that organizes similarity computation into correlation-consistent feature groups. This design improves numerical stability and preserves semantic coherence across feature subsets, thereby strengthening interpretability while mitigating instability in high-variance dimensions. The proposed fusion-score stage operates as a non-probabilistic decision rule that aggregates overlap-derived similarity channels into a per-class margin-like score and predicts via an $\arg\min$ criterion. As a lightweight and data-efficient classifier, this fusion mechanism is particularly suitable for small-to-moderate datasets, where stable primary decision rules are preferable to heavily parameterized models. For more complex regimes where direct fusion scoring may become suboptimal, we introduced a $\Delta$-distance construction that forms compact, contrastive class-referenced difference features. These $\Delta$-features preserve geometric separability while reducing dimensionality, enabling a more concise representation of discriminative information.

Building upon this compact representation, we integrated a variational quantum classifier (VQC) as an on-demand decision layer. Trained on $\Delta$-features, the VQC acts as a selective nonlinear discriminator for large-scale and highly imbalanced settings, while maintaining the geometry-driven front-end of the pipeline. Empirically, across Heart Disease, Breast Cancer, and Wine Quality datasets, the fusion-score classifier achieves competitive performance relative to classical baselines, consistent with its intended role as a stable and interpretable primary classifier under limited data conditions. For the Credit Card Fraud dataset, full-dataset evaluation underscores the importance of operating-point analysis—particularly alert rate and minority recall/precision trade-offs. In this extreme imbalance regime, the $\Delta$+VQC pipeline demonstrates the ability to recover high minority recall while maintaining a controlled alert budget. We further emphasize that operational evaluation is essential in rare-event detection. Default decision thresholds can misrepresent performance under extreme imbalance; therefore, alert-rate-matched comparisons are treated as the principal evaluation criterion in fraud detection scenarios, with fixed-threshold metrics serving only as secondary references. To support reproducibility and deployment-style evaluation, the complete pipeline is accompanied by practical artifacts, including fitted scalers, medoids, tuned fusion parameters, $\Delta$-feature construction procedures, and trained VQC checkpoints, enabling consistent scoring of new records. Looking forward, several directions merit further investigation. These include calibration under class-prior shift for the VQC stage; more principled regularization and stability control mechanisms for CGR configuration and fusion-weight tuning; hardware-executed overlap estimation and VQC training with noise-aware performance analysis; and broader benchmarking across diverse non-PCA tabular datasets, including scenarios with distribution drift and evolving subpopulation prototypes. Collectively, these extensions aim to strengthen the theoretical grounding, practical robustness, and real-world deployability of the proposed geometry-driven quantum-classical hybrid framework.

\section*{Acknowledgment}
The authors express gratitude to the IBM Quantum Experience platform and its team for creating the Qiskit platform and granting free access to their simulators for executing quantum circuits and conducting the experiments detailed below. The authors express appreciation for the Centre for Quantum Software and Information (CQSI) .

\section{Statements and Declarations}
\textbf{Competing Interests}: The authors have no financial or non-financial competing interests.\\
\textbf{Authors' contributions}:
The authors confirm their contribution to the paper as follows: 
Study conception and design: A.A.P., N.M., B.K.B., B.M.;\\ 
Data collection: A.A.P., N.M.;\\ 
Analysis and interpretation of results: A.A.P., N.M., B.K.B., B.M.;\\ 
Draft manuscript preparation: A.A.P., N.M., B.K.B.;\\
All authors reviewed the results and approved the final version of the manuscript.\\
\textbf{Funding}: Authors declare that there has been no external funding.\\
\textbf{Availability of data and materials}: All the data provided in this manuscript is generated during the simulation and can be provided upon reasonable request.

\bibliography{sn-article}% common bib file

\begin{thebibliography}{10}
\providecommand{\doi}[1]{\url{https://doi.org/#1}}
\bibcommenthead

\bibitem[\protect\citeauthoryear{He and Garcia}{2009}]{HeGarcia2009}
He H, Garcia EA.
\newblock Learning from Imbalanced Data.
\newblock IEEE Transactions on Knowledge and Data Engineering. 2009;21(9):1263--1284.
\newblock \doi{10.1109/TKDE.2008.239}.

\bibitem[\protect\citeauthoryear{Gama et~al.}{2014}]{Gama2014}
Gama J, {\v{Z}}liobait{\. e} I, Bifet A, Pechenizkiy M, Bouchachia A.
\newblock A Survey on Concept Drift Adaptation.
\newblock ACM Computing Surveys. 2014;46(4):44:1--44:37.
\newblock \doi{10.1145/2523813}.

\bibitem[\protect\citeauthoryear{Beyer et~al.}{1999}]{Beyer1999}
Beyer K, Goldstein J, Ramakrishnan R, Shaft U.
\newblock When Is ``Nearest Neighbor'' Meaningful?
\newblock In: Proceedings of the 7th International Conference on Database Theory (ICDT); 1999. .

\bibitem[\protect\citeauthoryear{Pestov}{1999}]{Pestov1999}
Pestov V.
\newblock Is the k-NN Classifier in High Dimensions Affected by the Curse of Dimensionality?
\newblock Computational Mathematics and Applications. 1999;.

\bibitem[\protect\citeauthoryear{Kaufman and Rousseeuw}{1987}]{KaufmanRousseeuw1987}
Kaufman L, Rousseeuw PJ.
\newblock Clustering by Means of Medoids.
\newblock In: Statistical Data Analysis Based on the L1 Norm and Related Methods; 1987. .

\bibitem[\protect\citeauthoryear{Buhrman et~al.}{2001}]{Buhrman2001}
Buhrman H, Cleve R, Watrous J, de~Wolf R.
\newblock Quantum Fingerprinting.
\newblock Physical Review Letters. 2001;87(16):167902.
\newblock \doi{10.1103/PhysRevLett.87.167902}.

\bibitem[\protect\citeauthoryear{Havl{\'i}{\v{c}}ek et~al.}{2019}]{Havlicek2019}
Havl{\'i}{\v{c}}ek V, C{\'o}rcoles AD, Temme K, Harrow AW, Kandala A, Chow JM, et~al.
\newblock Supervised Learning with Quantum-Enhanced Feature Spaces.
\newblock Nature. 2019;567:209--212.
\newblock \doi{10.1038/s41586-019-0980-2}.

\bibitem[\protect\citeauthoryear{Schuld and Killoran}{2019}]{SchuldKilloran2019}
Schuld M, Killoran N.
\newblock Quantum Machine Learning in Feature Hilbert Spaces.
\newblock Physical Review Letters. 2019;122(4):040504.
\newblock \doi{10.1103/PhysRevLett.122.040504}.

\bibitem[\protect\citeauthoryear{Chawla et~al.}{2002}]{Chawla2002}
Chawla NV, Bowyer KW, Hall LO, Kegelmeyer WP.
\newblock SMOTE: Synthetic Minority Over-sampling Technique.
\newblock Journal of Artificial Intelligence Research. 2002;16:321--357.

\bibitem[\protect\citeauthoryear{Wiebe et~al.}{2015}]{Wiebe2015}
Wiebe N, Kapoor A, Svore KM.
\newblock Quantum Algorithms for Nearest-Neighbor Methods for Supervised and Unsupervised Learning.
\newblock Quantum Information \& Computation. 2015;15(3--4):318--358.

\bibitem[\protect\citeauthoryear{Schuld et~al.}{2017}]{Schuld2017Distance}
Schuld M, Sinayskiy I, Petruccione F.
\newblock Implementing a Distance-Based Classifier with a Quantum Interference Circuit.
\newblock EPL (Europhysics Letters). 2017;119(6):60002.
\newblock \doi{10.1209/0295-5075/119/60002}.

\bibitem[\protect\citeauthoryear{Nielsen and Chuang}{2000}]{NielsenChuang2000}
Nielsen MA, Chuang IL.
\newblock Quantum Computation and Quantum Information.
\newblock Cambridge University Press; 2000.

\bibitem[\protect\citeauthoryear{Garc{\'i}a-Escart{\'i}n and Chamorro-Posada}{2013}]{GarciaEscartin2013}
Garc{\'i}a-Escart{\'i}n JC, Chamorro-Posada P.
\newblock The Swap Test and the Hong--Ou--Mandel Effect Are Equivalent.
\newblock Physical Review A. 2013;87(5):052330.
\newblock \doi{10.1103/PhysRevA.87.052330}.

\bibitem[\protect\citeauthoryear{Schuld and Petruccione}{2018}]{SchuldPetruccione2018}
Schuld M, Petruccione F.
\newblock Supervised Learning with Quantum Computers.
\newblock Springer; 2018.

\end{thebibliography}

\section{Appendix}
\subsection{VQC additional artifcats }
This section details the end-to-end VQC training and inference pipeline, including forward probability evaluation, SPSA-based optimization with minibatching and stabilization strategies, and artifact persistence for reproducible deployment. Together, these components ensure numerically stable training, consistent class decoding, and reliable inference aligned with the geometry-driven front-end of the framework.

\subsubsection{Forward probability evaluation and class decoding}
\label{subsec:vqc_forward}
We evaluate class probabilities from measurement outcomes. In our experiments we use statevector simulation to compute exact probabilities;in shot-based settings, the same procedure applies with empirical frequencies. Algorithm \ref{alg:vqc_forward_probs} computes class probability distributions by measuring the parameterized quantum circuit and decoding basis-state outcomes via direct or parity-based mapping.

\begin{algorithm}[H]
\caption{Forward pass: class probabilities from a parameterized VQC}
\label{alg:vqc_forward_probs}
\begin{algorithmic}[1]
\Require Circuit $U(\theta;x)$; inputs $A\in\mathbb{R}^{N\times m}$; number of classes $C$; mapping mode \texttt{direct} or \texttt{parity}; $n$ qubits.
\Ensure Probability matrix $P\in[0,1]^{N\times C}$.

\For{$i=1$ to $N$}
    \State Prepare state $\ket{\psi_i}=U(\theta;A_i)\ket{0}^{\otimes n}$.
    \State Obtain basis probabilities $p(s)$ for $s\in\{0,\dots,2^n-1\}$.
    \If{$C=2$ and \texttt{parity} mapping}
        \State $P_{i,0}\gets \sum_{s:\,\mathrm{parity}(s)=0} p(s)$;\quad $P_{i,1}\gets \sum_{s:\,\mathrm{parity}(s)=1} p(s)$.
    \Else \Comment{\texttt{direct} mapping}
        \State $P_{i,c}\gets p(c)$ for $c=0,\dots,C-1$ and renormalize over $c$.
    \EndIf
\EndFor
\State \Return $P$.
\end{algorithmic}
\end{algorithm}

\subsubsection{SPSA optimization with minibatching, gradient clipping, and early stopping}
\label{subsec:vqc_spsa}
We optimize VQC parameters using SPSA, requiring only two loss evaluations per iteration. Minibatching reduces computational cost in large datasets while preserving stable updates. Algorithm \ref{alg:vqc_spsa_minibatch} trains VQC parameters using stochastic perturbation-based gradient estimation with minibatching, gradient clipping, and early stopping for stable and efficient optimization.

\begin{algorithm}[H]
\caption{SPSA + minibatch VQC training (statevector forward)}
\label{alg:vqc_spsa_minibatch}
\begin{algorithmic}[1]
\Require Training angles $A^{(\mathrm{tr})}$ and labels $y^{(\mathrm{tr})}$; circuit builder Algorithm~\ref{alg:vqc_circuit_build}; reps $L$; steps $T$; batch size $B$; SPSA constants $(a,c,\alpha,\gamma)$; clip norm $G_{\max}$; early-stop patience $p$; tolerance $\delta$.
\Ensure Trained parameters $\theta^\star$ and training record.

\State Initialize $\theta\sim\mathcal{N}(0,\sigma^2)$.
\State $best\_loss \gets +\infty$;\quad $no\_improve\gets 0$.
\For{$t=1$ to $T$}
    \State $a_t \gets a/t^{\alpha}$;\quad $c_t \gets c/t^{\gamma}$.
    \State Sample minibatch indices $\mathcal{B}$ of size $B$ (without replacement).
    \State Draw Rademacher perturbation $\Delta\in\{-1,+1\}^{|\theta|}$.
    \State $\theta^{+}\gets \theta + c_t\Delta$;\quad $\theta^{-}\gets \theta - c_t\Delta$.
    \State Compute $f^{+}\gets \mathrm{CE}(\theta^{+};\mathcal{B})$ and $f^{-}\gets \mathrm{CE}(\theta^{-};\mathcal{B})$ using Algorithm~\ref{alg:vqc_forward_probs}.
    \State SPSA gradient estimate: $\widehat{g}\gets \frac{f^{+}-f^{-}}{2c_t}\cdot \Delta^{-1}$ (elementwise).
    \State Clip $\widehat{g}\gets \widehat{g}\cdot \min\{1, G_{\max}/\|\widehat{g}\|_2\}$.
    \State Update $\theta\gets \theta - a_t\widehat{g}$.
    \State $mb\_loss \gets (f^{+}+f^{-})/2$.
    \If{$mb\_loss < best\_loss - \delta$}
        \State $best\_loss\gets mb\_loss$;\quad $no\_improve\gets 0$.
    \Else
        \State $no\_improve\gets no\_improve+1$.
    \EndIf
    \If{$no\_improve \ge p$} \State \textbf{break} \EndIf
\EndFor
\State Compute final loss on an evaluation subset of training samples.
\State \Return $\theta^\star\leftarrow \theta$ and training record.
\end{algorithmic}
\end{algorithm}

\subsubsection{VQC artifact persistence and reproducible inference}
To ensure that VQC results are reproducible and that the trained model can be applied consistently to new records, we persist a complete artifact bundle after training. The saved bundle includes: (i) the finalized circuit structure and learned parameters $\theta^\star$, (ii) the preprocessing objects required to reconstruct the angle-encoded inputs (training-fitted scaler, clipping bound $z_{\max}$, and feature scaling coefficients $\lambda$), (iii) the decoding configuration (class mapping mode and, for binary tasks, the selected decision threshold $\tau$), and (iv) evaluation summaries (validation/test metrics and confusion matrices). This packaging mirrors the inference-time pipeline exactly and prevents train--test leakage by enforcing that all transforms are derived from training splits only. The persistence procedure is formalized in Algorithm~\ref{alg:vqc_save_artifacts}.
\label{app:vqc_artifacts}
\begin{algorithm}[H]
\caption{VQC artifact saving for reproducible scoring}
\label{alg:vqc_save_artifacts}
\begin{algorithmic}[1]
\Require Trained model $\{U(\theta^\star;x),\theta^\star\}$; scaler $\mathcal{S}$; feature list; clip bound $z_{\max}$; optional threshold $\tau$; metadata (k, hp, scores).
\Ensure Persisted artifact bundle.

\State Save circuit serialization (e.g., QPY); save $\theta^\star$ vector; save scaler object.
\State Write metadata JSON: feature columns, $z_{\max}$, $\tau$, $n$ qubits, reps, reupload, final loss, validation/test metrics.
\State Optionally write an alias pointer for the best model across $k$.
\end{algorithmic}
\end{algorithm}

\subsection{Fusion Score Evaluation of Test datasets}Additional tables for Fusion score evaluation.

\newpage

\begin{table*}
\centering
\caption{Dataset characteristics and Fusion-score performance (best-$k$ selected by coordinate descent; $\alpha$ tuned via $\alpha$-sweep on validation).}
\label{tab:fusion_benchmarks_summary}
\resizebox{\textwidth}{!}{
\begin{tabular}{lrrrrlrrrr}
\toprule
Dataset & $N$ & $d$ & \#classes & Class counts & Split (train/val/test) & best-$k$ & $\alpha^\star$ & Val Acc & Val F1$_{\text{macro}}$ \\
\midrule
Heart Disease & 918 & 11 & 2 & (0:410, 1:508) & 516/172/230 & 3 & 1.00 & 0.8198 & 0.8184 \\
Breast Cancer (Wisconsin) & 569 & 30 & 2 & (B:357, M:212) & 319/107/143 & 3 & 0.90 & 0.9159 & 0.9052 \\
Wine Quality (3-class) & 178 & 13 & 3 & (1:59, 2:71, 3:48) & 99/34/45 & 4 & 0.85 & 0.9412 & 0.9428 \\
\bottomrule
\end{tabular}
}\end{table*}

\begin{table*}
\centering
\caption{Test-set Fusion-score performance at the selected best-$k$ and $\alpha^\star$.}
\label{tab:fusion_benchmarks_test}
\begin{tabular}{lrr}
\toprule
Dataset & Test Acc & Test F1$_{\text{macro}}$ \\
\midrule
Heart Disease & 0.8478 & 0.8463 \\
Breast Cancer (Wisconsin) & 0.8881 & 0.8703 \\
Wine Quality (3-class) & 0.9556 & 0.9522 \\
\bottomrule
\end{tabular}
\end{table*}

\begin{table*}
\centering
\caption{Fusion-score confusion matrices (rows=true, cols=pred) for the selected best-$k$ per dataset.}
\label{tab:fusion_conf_mats}
\resizebox{\textwidth}{!}{
\begin{tabular}{llp{7.5cm}}
\toprule
Dataset & Split & Confusion matrix \\
\midrule
Heart Disease & Val  & $\begin{bmatrix} 63 & 14 \\ 17 & 78 \end{bmatrix}$ \\
Heart Disease & Test & $\begin{bmatrix} 86 & 17 \\ 18 & 109 \end{bmatrix}$ \\
\midrule
Breast Cancer (Wisconsin) & Val  & $\begin{bmatrix} 67 & 0 \\ 9 & 31 \end{bmatrix}$ \\
Breast Cancer (Wisconsin) & Test & $\begin{bmatrix} 90 & 0 \\ 16 & 37 \end{bmatrix}$ \\
\midrule
Wine Quality (3-class) & Val  & $\begin{bmatrix} 10 & 1 & 0 \\ 0 & 13 & 1 \\ 0 & 0 & 9 \end{bmatrix}$ \\
Wine Quality (3-class) & Test & $\begin{bmatrix} 15 & 0 & 0 \\ 0 & 18 & 0 \\ 0 & 2 & 10 \end{bmatrix}$ \\
\bottomrule
\end{tabular}
}
\end{table*}

\end{document}